\documentclass[reprint,showpacs,showkeys,preprintnumbers,amsmath,amssymb,aps,prb]{revtex4-2}

\usepackage{graphicx} % Include figure files
\usepackage{rotating}
\usepackage{color}
\usepackage{float}
\usepackage{multirow}
\begin{document}

\preprint{G. H. F.; Rh$_2$CoSb 2020}

\title{Phase transition in the magnetocrystalline anisotropy of tetragonal Heusler alloys: \\ 
       Rh$_2T$Sb, $T=$~Fe, Co}

\author{Gerhard H. Fecher}
\email{fecher@cpfs.mpg.de}
\affiliation{Max-Planck-Institute for Chemical Physics of Solids,
             D-01187 Dresden, Germany}

\author{Yangkun He}
\affiliation{Max-Planck-Institute for Chemical Physics of Solids,
             D-01187 Dresden, Germany}

\author{Claudia Felser}
\affiliation{Max-Planck-Institute for Chemical Physics of Solids,
             D-01187 Dresden, Germany}

\date{\today}

%%%%%%%%%%%%%%%%%%%%%%%%%%%%%%%%%%%%%%%%%%%%%%%%%%%%%%%%%%%%%%%%%%%%%%%%%%%%%%%%%
\begin{abstract}

This work reports on first principles calculations of the electronic and
magnetic structure of tetragonal Heusler compounds with the composition
Rh$_2$Fe$_{x}$Co$_{1-x}$Sb ($0\leq x\leq1$). It is found that the magnetic
moments increase from 2 to 3.4~$\mu_B$ and the Curie temperature decreases from
500 to 464~K with increasing Fe content $x$. The $3d$ transition metals make the
main contribution to the magnetic moments, whereas Rh contributes only
approximately 0.2~$\mu_B$ per atom, independent of the composition. The paper
focuses on the magnetocrystalline anisotropy of the borderline compounds
Rh$_2$FeSb, Rh$_2$Fe$_{0.5}$Co$_{0.5}$Sb, and Rh$_2$CoSb. A transition from
easy-axis to easy-plane anisotropy is observed when the composition changes from
Rh$_2$CoSb to Rh$_2$FeSb. The transition occurs at an iron concentration of
approximately 40\%.

\end{abstract}
%%%%%%%%%%%%%%%%%%%%%%%%%%%%%%%%%%%%%%%%%%%%%%%%%%%%%%%%%%%%%%%%%%%%%%%%%%%%%%%%%

%\pacs{75.30.-m, 71.20.Be, 61.18.Fs}
%PACS, the Physics and Astronomy Classification Scheme.

\keywords{Electronic structure, Magnetocrystalline anisotropy, Intermetallic compounds, Rh$_2$FeSb, Rh$_2$CoSb}

\maketitle

%%%%%%%%%%%%%%%%%%%%%%%%%%%%%%%%%%%%%%%%%%%%%%%%%%%%%%%%%%%%%%%%%%%%%%%%%%%%%%%%

\section{Introduction} %%%%%%%%%%%%%%%%%%%%%%%%%%%%%%%%%%%%%%%%%%%%%%%%%%%%%%%%%

Permanent or hard magnets are made of bulk materials with strong anisotropy,
which may be based on magnetocrystalline, shape anisotropy, or both. In magnets
with magnetocrystalline anisotropy, there should be only one easy crystal axis
of magnetisation so that the anisotropy is uniaxial. Such an uniaxial
magnetocrystalline anisotropy is found, for example, in tetragonal or hexagonal
systems.  Heusler alloys are compounds with formula $T_2T'M$, where $T$ and $T'$
are transition metals, and $M$ is a main group element. Some of these compounds
and alloys crystallise in tetragonal structure; however, most of them have a
cubic crystal structure. One advantage of Heusler compounds is that most of them
do not contain rare earth elements; rather, the magnetic properties are provided
by $3d$ transition metals. Many tetragonal Heusler alloys are Mn-based, and
several exhibit structural martensite--austenite phase transitions. In
particular, in the {\it inverse} structures with space group
$I\:\overline{4}m2$, the magnetic moments of the Mn atoms exhibit antiparallel
coupling. Thus, these alloys are generally ferrimagnets with low saturation
magnetisation. The Rh$_2TM$ alloys ($T'=$~V, Mn, Fe, Co; $M=$~Sn, Sb)
crystallise in a regular tetragonal structure with space group $I\:4/mmm$ and
are expected to exhibit uniaxial anisotropy when the $3d$ transition metals have
large moments.

Experiments on the crystal structure and magnetic properties of Rh$_2$-based
Heusler compounds were reported by Dhar {\it et al.}~\cite{DGM80}, who observed
a tetragonal structure and a magnetic moment of 1.4~$\mu_B$ in the primitive
cell. A Curie temperature of approximately 450~K was measured. Further, Fallev
{\it et al.} recently reported {\it ab initio} calculations for many tetragonal
Heusler compounds (including Rh$_2$FeSb and Rh$_2$CoSb)~\cite{FFJ17}. This work
proposed that thin films of Rh$_2$CoSb exhibit uniaxial, perpendicular
anisotropy with the easy direction along the $c$ ($[001]$) axis. Experiments and
calculations both suggest that Rh$_2$CoSb might be suitable hard magnetic
material with uniaxial anisotropy. However, the constituent elements, in
particular Rh, might be too expensive for applications where bulk materials are
needed, for example, permanent magnets in electric engines. However, the cost of
the materials is not as important for thin film applications, for example,
magnetic recording media or magnetoelectronic memory devices.

We recently reported experiments on the magnetic properties of Rh$_2$CoSb
\textcolor{red}{[in print, will be added later]}. It was found that Rh$_2$CoSb
has uniaxial anisotropy, where $c$ is the {\it easy} axis. The present work
describes theoretically the magnetic properties of Rh$_2$CoSb, its sister
compound Rh$_2$FeSb, and alloys with mixed Co$_{1-x}$Fe$_x$ composition.

\section{Details of the calculations} %%%%%%%%%%%%%%%%%%%%%%%%%%%%%%%%%%%%%%%%%

The electronic and magnetic structures of Rh$_2T$Sb ($T=$~Fe, Co) were
calculated using {\scshape Wien}2k~\cite{BSS90,SBl02,BSM01} and {\scshape
Sprkkr}~\cite{Ebe99,EKM11} in the local spin density approximation. In
particular, the generalised gradient approximation of Perdew, Burke, and
Ernzerhof~\cite{PBE96} was used to parametrise the exchange correlation
functional. A $k$-mesh based on $126\times126\times126$ points of the full
Brillouin zone was used for integration when the total energies were calculated
to determine the magnetocrystalline anisotropy (see also
Appendix~\ref{app:app1}). The calculations are described in greater detail in
References~\cite{KFF07b,FCF13}. The spin spirals and magnons were calculated
according to the schemes described in References~\cite{MFE11} and~\cite{TCF09},
respectively. Calculations for the disordered or off-stoichiometric compounds
with mixed site occupations were performed using {\sc Sprkkr} and the coherent
potential approximation (CPA)~\cite{Sov67} in the full potential mode. The CPA
allows the simulation of random site occupation by different elements.
Complications arising in the calculation of the magnetic anisotropy energies are
discussed in detail by Khan {\it et al.}~\cite{KBE16}, who compared results
obtained using {\scshape Wien}2k and {\scshape SPRKKR}.

\begin{figure}[htb]
   \centering
   \includegraphics[width=8cm]{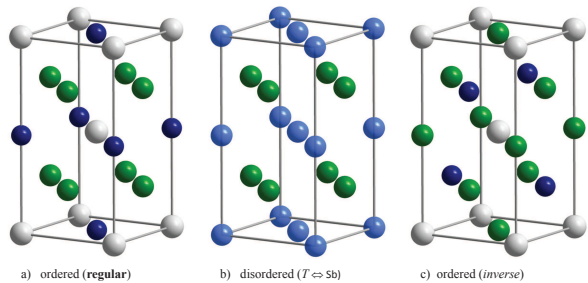}
   \caption{Crystal structure of Rh$_2T$Sb ($T=$~Fe, Co). \newline
            In the well-ordered regular structure (a), the sites of the lattice with space group
            $I\:4/mmm$ (139) are occupied as follows:
            4d (0 1/2 1/4), Rh; 2b (0 0 1/2), $T$; and 2a (0 0 0), Sb.
            In the disordered structure (b), the $T$ and Sb atoms are randomly distributed
            on the 2b and 2a sites. The {\it inverse} tetragonal structure is shown in (c) for
            comparison.}
   \label{fig:struct}
\end{figure}

The basic crystal structure of the tetragonal Heusler compounds [prototype,
Rh$_2$VSn; $tI8$; $I\:4/mmm$ (139) $dba$] is shown in
Figure~\ref{fig:struct}(a). The atoms are located in the ferromagnetic structure
on the 4d, 2b, and 2a Wyckoff positions of the centred tetragonal cell. The
magnetic order changes the symmetry, and the resulting magnetic space group for
collinear ferromagnetic order with moments along the $c$ axis is $I\:4/mm'm'$
(139.537), where $'$ is the spin reversal operator~\cite{Jos91}. The symmetry is
reduced to that of space group $I\:m'm'm$ (71.536) when the magnetisation
$\vec{M}$ is along the $a$ axis ($[100]$) or $F\:m'm'm$ (69.524) for
$\vec{M}\|[110]$.

%%%%%%%%%%%%%%%%%%%%%%%%%%%%%%%%%
%%%%%%%%%%%%%%%%%%%%%%%%%%%%%%%%%%%%%%%

The electronic structure and magnetic properties were calculated using the
optimised lattice parameters. As starting point, the lattice parameters of two
alternative structures were optimised using {\scshape Wien}2k. In addition to
the regular Heusler structure described above, the inverse structure with space
group $I\:\overline{4}m2$ (119) $dbca$ was assumed. In this structure, the
positions of the Co atom and one of the Rh atoms are interchanged. Spin--orbit
interaction was considered owing to the high $Z$ values of Rh and Sb. Note that
the spin--orbit interaction is an intrinsic property in the fully relativistic
{\scshape Sprkkr} calculations, which solve the Dirac equation. The results of
the optimisation are summarised in Table~\ref{tab:opt}. The regular structure is
found to have lower energy; it thus describes the ground state. The energy
difference compared to the inverse structure is approximately 430~meV. The
formation enthalpy is calculated as
$\Delta H_{f}=E_{tot}-(2E_{\rm Rh}+E_{\rm Co}+E_{\rm Sb})$,
that is, the difference between the total energy of the compound in different
structures and the sum of the energies of the elements in their ground state
structure. The formation enthalpy is clearly lower for the regular structure
than for the inverse tetragonal structure. Note that the formation enthalpy is
even lower (-220 meV) for the cubic $L2_1$ structure. The calculated lattice
parameters are in good agreement with experimental values~\cite{DGM80}; however,
the calculated $c$ value and $c/a$ ratio are approximately 4\% larger. This
finding might be explained by either a temperature effect or some disorder in
the experiment.

% Table N %%%%%%%%%%%%%%%%%%%%%%%%%%%%%%%%%%%%%%%%%%%%%%%%%%%%%%%%%%%%%%%%%%%%%%%
\begin{table}[htb]
\centering
    \caption{Structural properties of Rh$_2$CoSb. \\
             Calculations are performed for the regular (139) and inverse (119) Heusler structures.
             The lattice parameters ($a$, $c$, $c/a$),
             formation enthalpy ($\Delta H_{f}$), and
             spin magnetic moment $m_{\rm s}$ of the primitive cell (total experimental magnetic moment) are listed.
             Experimental values from Reference~[\onlinecite{DGM80}] are shown for comparison.
             Note that the magnetic moment in this reference is not saturated. }
    \begin{ruledtabular}
    \begin{tabular}{l cccc}
                              & \multicolumn{2}{c}{Calculated} & \multicolumn{2}{c}{Exp.} \\
                             & 139     &  119     & here   & [\onlinecite{DGM80}] \\
       \hline
       $a$   [\AA]            & 4.0104  &  3.95    & 4.0393 & 4.04  \\
       $c$   [\AA]            & 7.3628  &  7.56    & 7.1052 & 7.08  \\
       $c/a$                  & 1.836   &  1.91    & 1.759  & 1.75  \\
       \hline
       $\Delta H_{f}$ [meV]   & -754    & -325     &        &       \\
       $m_{\rm s}$ [$\mu_B$]  & 2.04    &  1.79    & 2.36   & 1.4   \\
       $T_C$ [K]              &         &          & 450    & 450   \\
    \end{tabular}
    \end{ruledtabular}
    \label{tab:opt}
\end{table}
%%%%%%%%%%%%%%%%%%%%%%%%%%%%%%%%%%%%%%%%%%%%%%%%%%%%%%%%%%%%%%%%%%%%%%%%%%%%%%%%%

\section{Results and discussion}

\subsection{Electronic and magnetic structure of Rh$_2$CoSb}

\begin{figure}[htb]
   \centering
   \includegraphics[width=8cm]{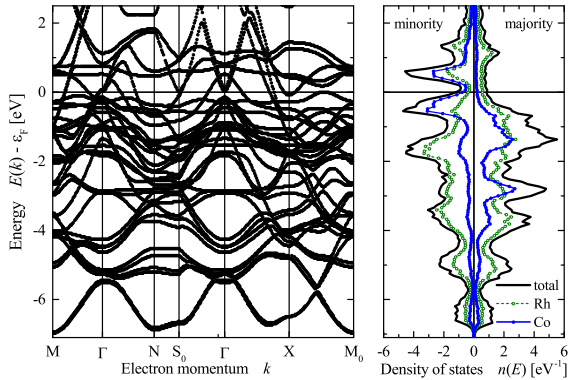}
   \caption{Electronic structure of Rh$_2$CoSb (I).\\
            Shown is the fully relativistic band structure together with the
            total and site (Rh and Co) specific, spin-resolved densities of states.
            }
   \label{fig:banddosCo}
\end{figure}

The calculated electronic structure of Rh$_2$CoSb in the regular tetragonal
Heusler structure is illustrated in Figure~\ref{fig:banddosCo} in terms of the band
structure and density of states  ($n(E)$). The relativistic bands,
spin-resolved total density of states, and its atomic contributions are shown. The
electronic structure is calculated in the full relativistic mode by solving the
Dirac equation. The band structure from semi-relativistic calculations is shown
in the Appendix.

%%%%%%%%%%%%%%%%%%%%%%%%%%%%%%%%%
%%%%%%%%%%%%%%%%%%%%%%%%%%%%%%%%%%%%%%%

Both rhodium and cobalt contribute to the magnetic moment of the compound. The
spin and orbital magnetic moments are
$m_s^{\rm Co}=1.656\:\mu_B$ and $m_l^{\rm Co}=0.139\:\mu_B$ for cobalt and
$m_s^{\rm Rh}=\:0.206\mu_B$ and $m_l^{\rm Rh}=0.007\:\mu_B$ for rhodium,
respectively. The overall magnetic moment (spin plus orbital) of the primitive
cell is
$m_{\rm tot}=2.188\:\mu_B$.
The orbital moment of the Co atoms makes a remarkably large contribution.

\begin{figure}[htb]
   \centering
   \includegraphics[width=8cm]{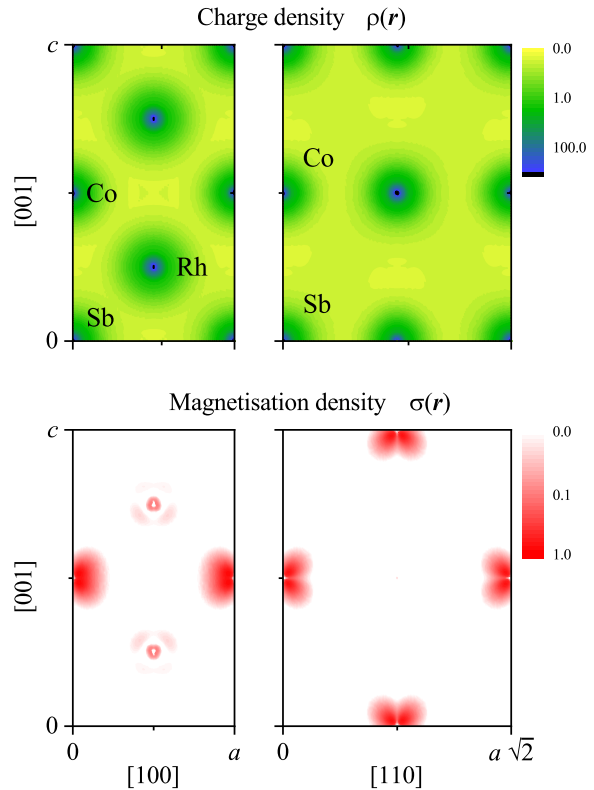}
   \caption{Electronic structure of Rh$_2$CoSb (II).\\
            Fully relativistic charge ($\rho(r)$) and spin ($\sigma(r)$) distributions
            for the (001) and (110) planes are shown. The calculation is for $m\|c$, that is,
            the magnetisation points along [001]
            according to the {\it easy axis} behaviour of the magnetic anisotropy.
            (Note: colour bars are in atomic units.) }
   \label{fig:chargeCo}
\end{figure}

The real space charge and spin distributions are shown in
Figure~\ref{fig:chargeCo}. The charge density ($\sigma(r)$) of the atoms has no
striking shape. It appears to be nearly spherical but still reflects the two- or
fourfold symmetry. As expected, most of the electrons are close to the ion
cores. By contrast, the spin or magnetisation density ($\sigma(r)$) has a much
more pronounced shape depending on the plane. In particular, in the (110) plane,
it has a distinct butterfly shape. The spin density is positive at both the Co
and Rh atoms. It is clearly higher near the Co atoms than near the Rh atoms,
which ultimately gives Co a higher magnetic moment. The magnetisation density of
the Rh atoms is aligned along the magnetisation direction and points somewhat
toward the nearest Co atoms.

%%%%%%%%%%%%%%%%%%%%%%%%%%%%%%%%%
%%%%%%%%%%%%%%%%%%%%%%%%%%%%%%%%%%%%%%%

\subsubsection{Magnetic anisotropy}

Further, the directional dependence of the magnetisation was investigated to
explain the collinear magnetic order in detail. In particular, the total energy
was calculated for cases where the magnetisation points along different
crystallographic directions. The obtained energy differences make it possible to
determine the magnetocrystalline anisotropy (see also Appendix~\ref{app:app1}).

In the magnetic anisotropy of Rh$_2$CoSb, the {\it easy} axis is along the $c$
($[001]$) axis. The simple second-order uniaxial anisotropy constant is
$K_u=1.37$~MJ/m$^3$ (see Equations~(\ref{eq:secndord}) and~(\ref{eq:simple}) in
Appendix~\ref{app:app1_uni}). This results in an anisotropy field of $\mu_0
H_u\approx2.4$~T. A more detailed analysis reveals that the simple second-order
anisotropy constant $K_u$ is not sufficient to describe the magnetocrystalline
anisotropy, as discussed in Section~\ref{sec:mca}.

Further, the dipolar magnetocrystalline anisotropy was calculated as described
in Appendix~\ref{app:app1_dipani} and was found to be
$\Delta E_{\rm dipaniso}=0.09$~$\mu$eV. The positive value indicates an easy
dipolar direction along the $[001]$ axis. The dipolar anisotropy is rather small
compared to the anisotropy calculated from the total energy. Here, it was
calculated for a sphere with a radius of 30~nm. The results for other shapes
will be different, resulting in a distinct shape anisotropy. In particular, in
thin films, the dimension perpendicular to the film is much smaller than the
dimensions in the film plane. Therefore, the summation in
Equation~(\ref{eq:dipaniso}) becomes a truncated sphere that is strongly
anisotropic, and a pronounced thin film anisotropy appears. This thin film
anisotropy will also be affected by the magnetic moments, which are different at
interfaces and surfaces from that at the centre layers of the film.

\subsubsection{Spiral spin order}

\begin{figure}[htb]
   \centering
   \includegraphics[width=8cm]{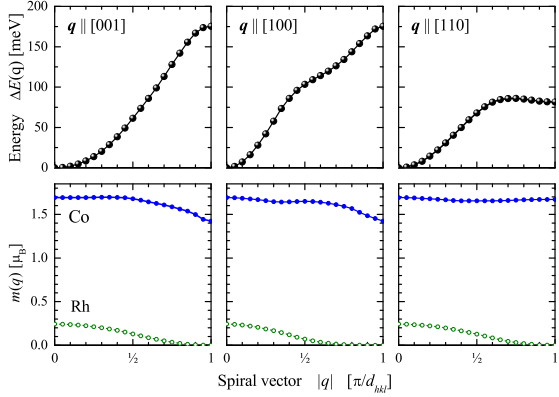}
   \caption{Planar spin spirals in Rh$_2$CoSb.\\
            The spiral energies are given with respect to $q=0$, that is, $\Delta E(q)=E(q)-E(0)$.
            The behaviour of the magnetic moment $m(q)$ is shown for Rh and Co.}
   \label{fig:spiral_comp}
\end{figure}

The energy of the spin spirals was calculated to search for non-collinear
magnetic order. The spin spirals were calculated for different directions and
different cones. In planar spirals, the spins are perpendicular to the
propagation direction. Figure~\ref{fig:spiral_comp} compares the energies of
planar spirals along the high-symmetry directions.

%%%%%%%%%%%%%%%%%%%%%%%%%%%%%%%%%
%%%%%%%%%%%%%%%%%%%%%%%%%%%%%%%%%%%%%%%

The spirals along $[100]$ or $[110]$ propagate in the fourfold plane, whereas
the spiral in the $[001]$ direction propagates along the $c$ axis. In all cases,
the lowest energy is observed at $q=0$. The magnetic moment of the Co atoms
varies by approximately 17\% at maximum. The magnetic moment of the Rh atoms
decreases with increasing $q$ and vanishes at the border, independent of the
propagation direction of the spiral.

\begin{figure}[htb]
   \centering
   \includegraphics[width=8cm]{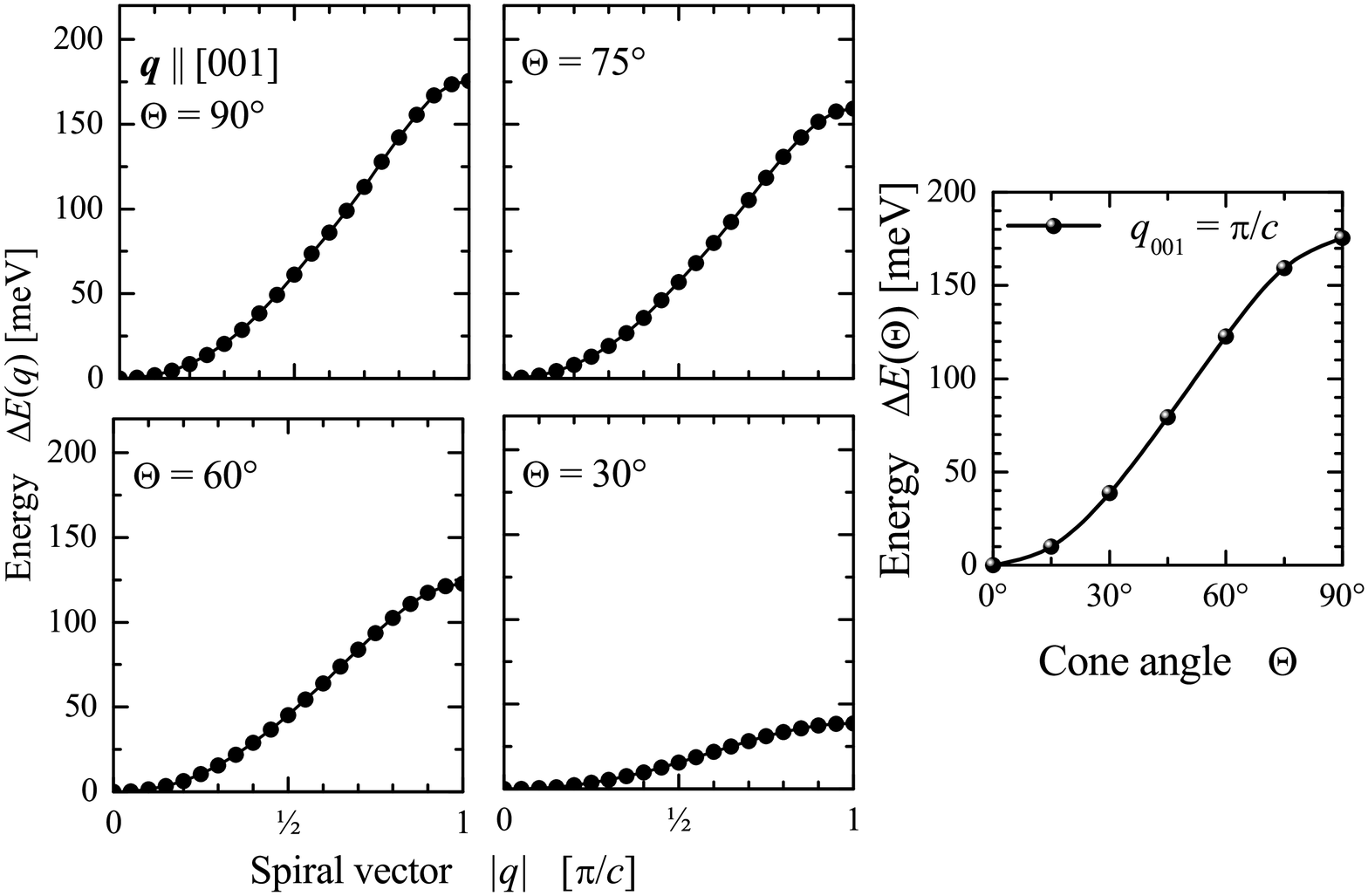}
   \caption{Conical spirals in Rh$_2$CoSb.\\
            Spiral energies for different cone angles
            and the wave vector along the $c$ axis ($q \| [001]$) are shown.
            The angular dependence at $q = \pi /c$ is also shown. }
   \label{fig:cone-spiral}
\end{figure}

The spin direction was assumed to be perpendicular to the $q$ vector in the
above calculations for planar spirals. Thus, the angle between $\vec{q}$ and the
local magnetic moment $\vec{m}_i$ was set to $\Theta=\pi/2$. Next, the spirals
were assumed to be conical with $0<\Theta<\pi/2$ to allow for a more detailed
analysis. The calculations were performed for $q$ along $[001]$.
Figure~\ref{fig:cone-spiral} displays the results for conical spirals with
various cone angles. The highest energies appear for the planar spiral. The
energy at the border of the Brillouin zone ($q=\pi/c$) exhibits a sine
dependence. Thus, it vanishes in the antiferromagnetic state. The behaviour of
the local magnetic moments suggests more localised behaviour at the Co atoms and
induced behaviour at the Rh atoms.

%%%%%%%%%%%%%%%%%%%%%%%%%%%%%%%%%
%%%%%%%%%%%%%%%%%%%%%%%%%%%%%%%%%%%%%%%

The calculated spiral energies indicate that this type of magnetic order is
rather improbable. The spiral energies increase monotonously with the wave
vector and cone angle, rather independent on the $\vec{q}$ direction. The
monotonic behaviour suggests that a canted magnetic order is also very
unlikely~\cite{LNB04}.

%%%%%%%%%%%%%%%%%%%%%%%%%%%%%%%%%%%%%%%%%%%%%%%%%%%%%%%%%%%%%%%%%%%%%%%%%%%%%%%%
\subsubsection{Exchange coupling and magnons}

\begin{figure}[htb]
   \centering
   \includegraphics[width=8cm]{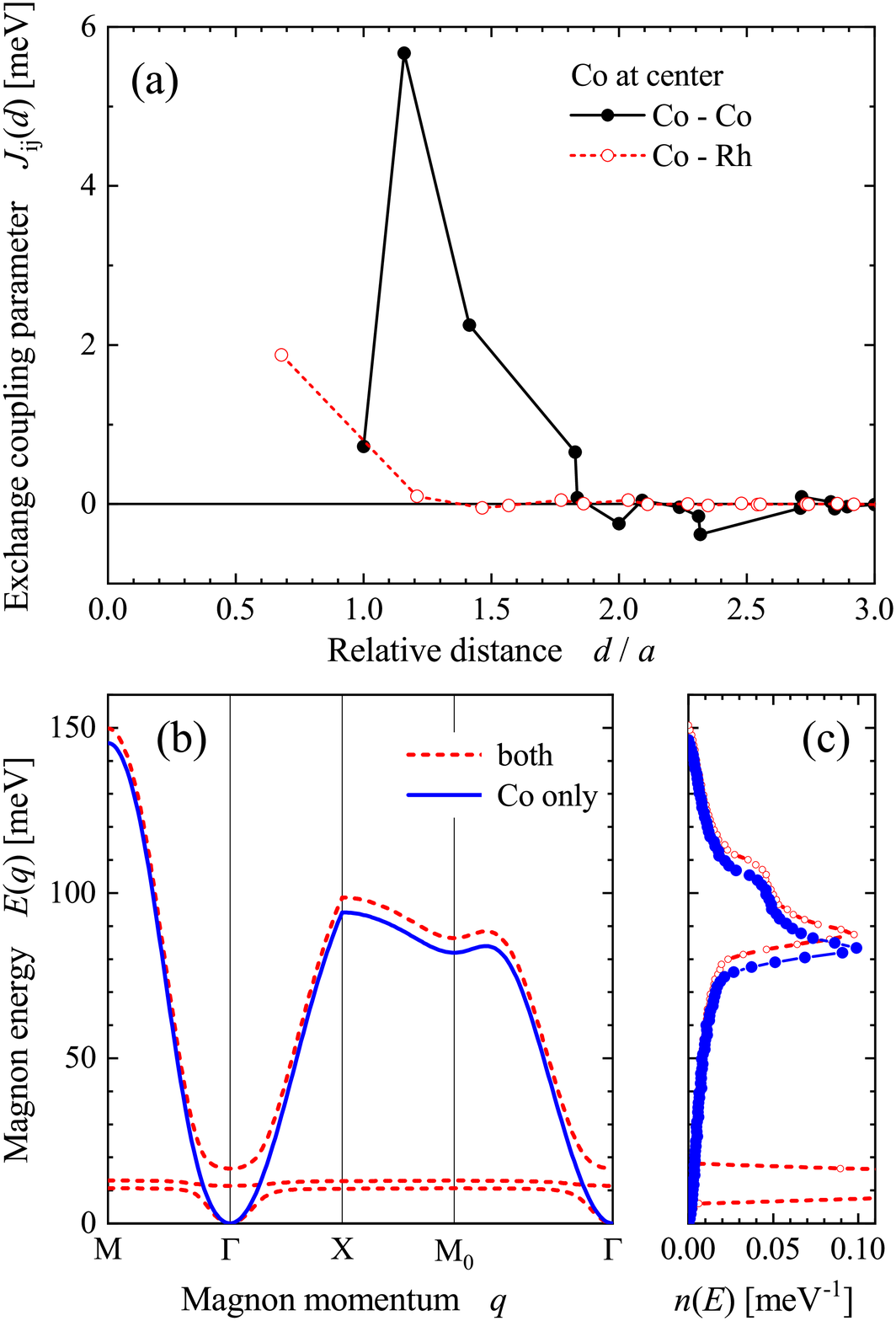}
   \caption{Exchange coupling parameters of Rh$_2$CoSb.\\
            (a) Exchange coupling parameters for Co--Co
            and Co--Rh interaction as functions of distance.
            Lines are drawn for better comparison.
            (b) Magnon dispersion and (c) density of states.
            Calculations were performed with and without
            accounting for the magnetic moment of the Rh atoms (that is, for both atoms and for Co only).}
   \label{fig:jxcmagnon}
\end{figure}

The exchange coupling energies were calculated using the scheme of Liechtenstein
{\it et al.}~\cite{LKG84,LKA87} to estimate the Curie temperature, spin
stiffness, and presence of magnons~\cite{TCF09}. The exchange coupling
parameters are plotted in Figure~\ref{fig:jxcmagnon}(a). The most dominant
parameters for Co--Co and Co--Rh interactions are shown; all the others are
comparatively small. The largest interaction appears for Co atoms in the centre
and nearest to the Co in the neighbouring plane. From the calculated exchange
coupling energies, the Curie temperature was found to be $T_C=498$~K, which is
close to the experimental value (450~K)~\cite{DGM80}. The calculated spin wave
stiffness constant is $D_{ij}=866$~meV$\cdot$\AA$^2$, and the interpolation
scheme of Padja {\it et al.}~\cite{PKT01} yields an extrapolated spin wave
stiffness of $D_0=864$~meV$\cdot$\AA$^2$.

%%%%%%%%%%%%%%%%%%%%%%%%%%%%%%%%%
%%%%%%%%%%%%%%%%%%%%%%%%%%%%%%%%%%%%%%%

The magnon dispersion was calculated by Fourier transformation of the real space
exchange coupling parameters. The result is presented in
Figure~\ref{fig:jxcmagnon}(b), and the magnon density of states is shown in
Figure~\ref{fig:jxcmagnon}(c). Two calculations were made; in one calculation,
only the Co--Co interaction was considered, and in the other, the moments of the
Rh atoms, which result in additional Co--Rh and Rh--Rh coupling, were included.
The latter calculation yields flat dispersion curves and a high density of
states. A comparison of the two calculations reveals that the magnons are
dominated by the Co--Co interaction. Note that the Curie temperature in only
10~K lower when the Rh moments and the corresponding exchange parameters are
ignored.

%%%%%%%%%%%%%%%%%%%%%%%%%%%%%%%%%%%%%%%%%%%%%%%%%%%%%%%%%%%%%%%%%%%%%%%%%%%%%%%%
\subsection{Results for Rh$_2$FeSb}
\label{sec:Fe}

The calculations for Rh$_2$FeSb were performed in the same way as for
Rh$_2$CoSb. The regular structure with space group no.~139 was found to be
more stable than the inverse structure with space group no.~119. In addition, as in
the case of Rh$_2$CoSb, the calculated $c$ lattice parameter, and thus $c/a$, are considerably
larger than the experimental values (see Table~\ref{tab:optFe}).

% Table N %%%%%%%%%%%%%%%%%%%%%%%%%%%%%%%%%%%%%%%%%%%%%%%%%%%%%%%%%%%%%%%%%%%%%%%
\begin{table}[htb]
\centering
    \caption{Structural properties of Rh$_2$FeSb. \\
             Calculations are performed for the regular tetragonal Heusler structures.
             Lattice parameters ($a$, $c$, $c/a$) and
             spin magnetic moment $m_{\rm s}$ of the primitive cell are listed.
             Experimental values from Reference~[\onlinecite{DGM80}] are shown for comparison.
             Note that the experimental moments in~[\onlinecite{DGM80}] are not saturated.}
    \begin{ruledtabular}
    \begin{tabular}{l ccc}
                              &            & \multicolumn{2}{c}{Experiment} \\
                              & Calculated &  This work & Ref. [\onlinecite{DGM80}] \\
       \hline
       $a$   [\AA]            & 4.0418     &  4.0671    & 4.07 \\
       $c$   [\AA]            & 7.3995     &  7.0161    & 6.96 \\
       $c/a$                  & 1.8308     &  1.7251    & 1.71 \\
       \hline
       $m_{\rm s}$ [$\mu_B$]  & 3.4        &  3.8       & 2.8 \\
       $T_C$ [K]              &            &  510       & 510 \\
    \end{tabular}
    \end{ruledtabular}
    \label{tab:optFe}
\end{table}
%%%%%%%%%%%%%%%%%%%%%%%%%%%%%%%%%%%%%%%%%%%%%%%%%%%%%%%%%%%%%%%%%%%%%%%%%%%%%%%%%

\begin{figure}[htb]
   \centering
   \includegraphics[width=8cm]{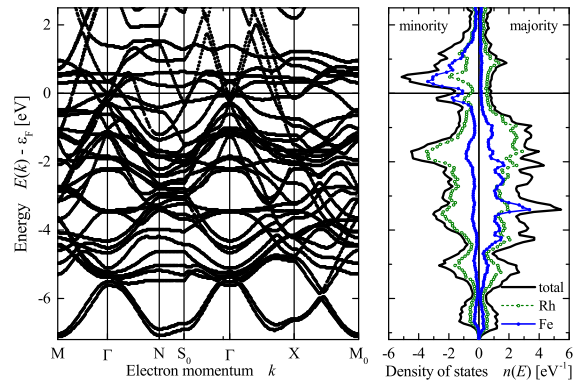}
   \caption{Electronic structure of Rh$_2$FeSb (I).\\
            Fully relativistic band structure is shown, along with the
            total- and site-specific spin-resolved densities of states for Rh and Fe. }
   \label{fig:banddosFe}
\end{figure}

The electronic structure of Rh$_2$FeSb is illustrated in
Figure~\ref{fig:banddosFe}. The fully relativistic band structure and the spin-
and site-resolved densities of states are shown. The calculated spin and orbital
magnetic moments are $m_s^{\rm Fe}=2.978\:\mu_B$ and $m_l^{\rm Fe}=0.080\:\mu_B$
for iron and $m_s^{\rm Rh}=\:0.228\mu_B$ and $m_l^{\rm Rh}=0.006\:\mu_B$ for
rhodium, respectively. The overall magnetic moment (spin plus orbital) of the
primitive cell is $m_{\rm tot}=3.488\:\mu_B$. The magnetic moment of the Fe
atoms is strongly localised, which is typical of Heusler compounds with high
magnetic moments. It clearly exceeds the value for elemental iron.

%%%%%%%%%%%%%%%%%%%%%%%%%%%%%%%%%
%%%%%%%%%%%%%%%%%%%%%%%%%%%%%%%%%%%%%%%

\begin{figure}[htb]
   \centering
   \includegraphics[width=8cm]{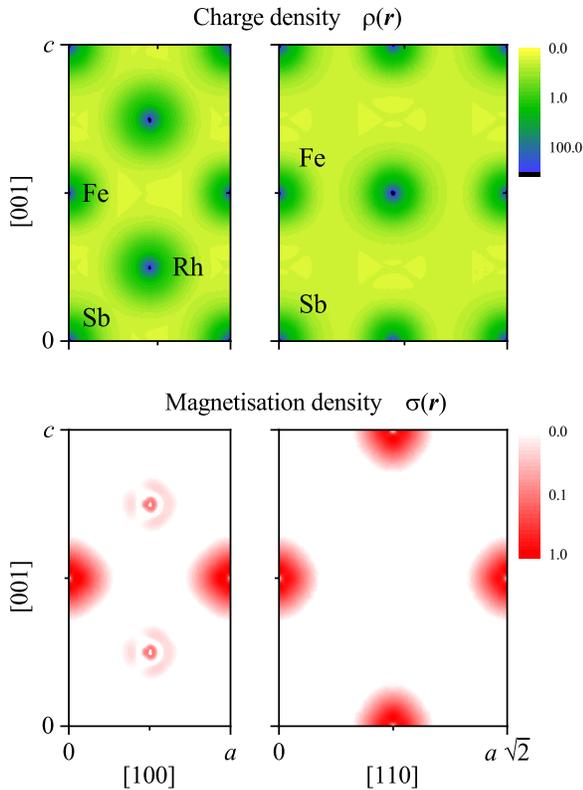}
   \caption{Electronic structure of Rh$_2$FeSb (II).\\
            Fully relativistic charge ($\rho(r)$) and spin ($\sigma(r)$) distributions
            for different planes. Magnetisation is perpendicular to $c$ with $m$ along [100]
            in accordance with the {\it easy plane} behaviour of the magnetic anisotropy.
            (Note: colour bars are in atomic units.) }
   \label{fig:chargeFe}
\end{figure}

The real space charge and spin distributions of Rh$_2$FeSb are shown in
Figure~\ref{fig:chargeFe}. As in the Co-containing compound, $\sigma(r)$ does
not have a pronounced shape (compare Figure~\ref{fig:chargeCo}). The
magnetisation density ($\sigma(r)$) around the Fe atoms has a less distinct
shape compared to Co in Rh$_2$CoSb; it is also not greatly affected by changes
in the magnetisation direction. The main difference is the magnetisation density
around the Rh atoms, which is rotated and appears to be aligned along the
magnetisation direction.

%%%%%%%%%%%%%%%%%%%%%%%%%%%%%%%%%
%%%%%%%%%%%%%%%%%%%%%%%%%%%%%%%%%%%%%%%

Table~\ref{tab:compCoFe} compares the calculated magnetic data of Rh$_2$CoSb and
Rh$_2$FeSb. Rh$_2$FeSb clearly has a smaller orbital magnetic moment than
Rh$_2$CoSb, whereas its spin magnetic moment is higher because of the effect of
the Fe atoms. The induced magnetic moments of the Rh atoms are similar in both
compounds.

% Table N %%%%%%%%%%%%%%%%%%%%%%%%%%%%%%%%%%%%%%%%%%%%%%%%%%%%%%%%%%%%%%%%%%%%%%%
\begin{table}[htb]
\centering
    \caption{Calculated magnetic properties of Rh$_2$FeSb, Rh$_2$Fe$_{0.5}$Co$_{0.5}$Sb, and Rh$_2$CoSb. \\
             Spin $m_s$ and orbital $m_l$ magnetic moments per atom
             (Rh, $T$ = Co, Fe with $m\|c$ in all cases)
             of the primitive cell ($total$) are listed, as well as Curie temperature $T_C$, spin stiffness $D_0$,
             and anisotropy parameters.
             (Note that the dipolar anisotropy is three orders of magnitude
             lower than the magnetocrystalline part.)}
    \begin{ruledtabular}
    \begin{tabular}{l ccc}
                                            & Fe    & Fe$_{0.5}$Co$_{0.5}$ & Co    \\
       \hline
       $m_s^{\rm Rh}$ [$\mu_B$]             & 0.237 &  0.239               & 0.204 \\
       $m_l^{\rm Rh}$ [$\mu_B$]             & 0.006 &  0.008               & 0.006 \\
       $m_s^{\rm Fe}$ [$\mu_B$]             & 3.006 &  2.977               & -     \\
       $m_l^{\rm Fe}$ [$\mu_B$]             & 0.080 &  0.084               & -     \\
       $m_s^{\rm Co}$ [$\mu_B$]             & -     &  1.747               & 1.674 \\
       $m_l^{\rm Co}$ [$\mu_B$]             & -     &  0.132               & 0.137 \\
      \hline
       $m_s^{total}$ [$\mu_B$]              & 3.44  &  2.81                & 2.04  \\
       $m_l^{total}$ [$\mu_B$]              & 0.09  &  0.12                & 0.15  \\
      \hline
       $T_C$ [K]                            & 465   &  480                 & 500   \\
       $D_0$ [meV~\AA$^2$]                  & 590   &  700                 & 870   \\
      \hline
       $K_u$ [MJ/m$^3$]                     & -1.21 &  -0.23               &  1.37 \\
       $|\mu_0 H_a|$ [T]                    &  1.34 &   0.31               &  2.43 \\
       $\Delta E_{\rm dipaniso}$ [kJ/m$^3$] &  1.9  &                      &  2.0  \\
    \end{tabular}
    \end{ruledtabular}
    \label{tab:compCoFe}
\end{table}
%%%%%%%%%%%%%%%%%%%%%%%%%%%%%%%%%%%%%%%%%%%%%%%%%%%%%%%%%%%%%%%%%%%%%%%%%%%%%%%%%

The calculated Curie temperatures are of the same order of magnitude as the
experimental values. In contrast to the calculated results, however, the
experimental value of the Fe compound is higher than that of the Co compound. A
possible reason is differences in the variation of the lattice parameters with
temperature, which affect the exchange coupling parameters and thus $T_C$ and
also the spin stiffness. Note that a much lower Curie temperature is obtained
for the Co compound when it is off-stoichiometric (see
Appendix~\ref{sec:offst}), whereas the $T_C$ value of the off-stoichiometric Fe
compound is slightly higher.

The anisotropy has the {\it hard} axis along the $z$ ($[001]$) direction, and
the easy plane is the the basal plane. By contrast, for Rh$_2$CoSb, the $z$
direction is the {\it easy} axis. The simple uniaxial anisotropy constant is
$K_u=- 1.21$~MJ/m$^3$. Consequently, the anisotropy field is
$|\mu_0 H_a|=1.34$~T. The appearance of the {\it "hard"} axis along $z$ is
opposite to Rh$_2$CoSb where $z$ is the {\it "easy"} axis. The dipolar
magnetocrystalline anisotropy of Rh$_2$FeSb is
$\Delta E_{\rm dipaniso}=0.09$~$\mu$eV, indicating that the easy dipolar
direction is along the $[001]$ axis, like that of the Co- containing compound.
This behaviour is caused by the strong magnetic moments of the $3d$ transition
metals, in addition to the elongation of the tetragonal crystal structure along
the $c$ axis.

\begin{figure}[htb]
   \centering
   \includegraphics[width=8cm]{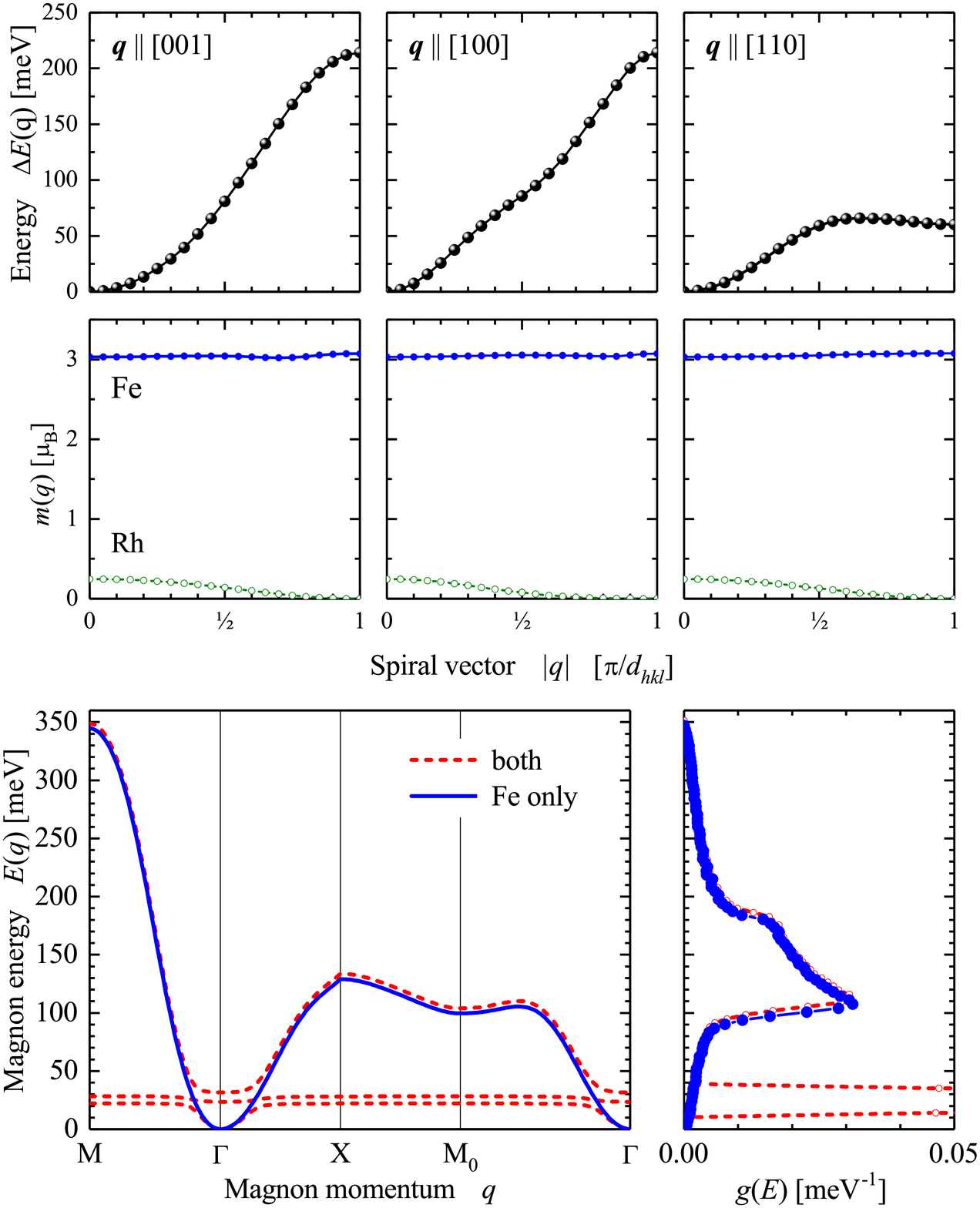}
   \caption{Dynamic magnetic properties of Rh$_2$FeSb.\\
            Spiral energies with corresponding magnetic moments
            and the magnon dispersion are shown, along with
            the magnon density of states $g(E)$.
            Magnon calculations were performed with and without the magnetic moment of the Rh atoms;
            in the latter case, all Fe--Rh interactions are neglected.}
   \label{fig:magnonFe}
\end{figure}

The dynamic magnetic properties of Rh$_2$FeSb are shown in
Figure~\ref{fig:magnonFe}. The spin spirals and magnons are similar to those of
Rh$_2$CoSb; however, their energies extend to higher values. The behaviour of
the spin spirals rules out the presence of non-collinear magnetic
structure~\cite{LNB04}.

%%%%%%%%%%%%%%%%%%%%%%%%%%%%%%%%%
%%%%%%%%%%%%%%%%%%%%%%%%%%%%%%%%%%%%%%%

%%%%%%%%%%%%%%%%%%%%%%%%%%%%%%%%%%%%%%%%%%%%%%%%%%%%%%%%%%%%%%%%%%%%%%%%%%%%%%%
\subsection{ Results for Rh$_2$Fe$_{x}$Co$_{1-x}$Sb }
\label{sec:FeCo}

Owing to the differences in magnetic anisotropy between the Fe- and Co-based
compounds, it is interesting to investigate a mixed system containing both Fe
and Co. Therefore, calculations were also performed for
Rh$_2$Fe$_{x}$Co$_{1- x}$Sb using {\sc Sprkkr} and the CPA. The CPA enables the
simulation of random occupation of Fe and Co atoms at a single site (here $2b$).
The obtained magnetic properties of Rh$_2$Fe$_{0.5}$Co$_{0.5}$Sb are shown in
Table~\ref{tab:compCoFe}. The uniaxial anisotropy constant is negative, like
that of Rh$_2$FeSb; however, its absolute value is considerably lower (by a
factor of 35) than that of Rh$_2$CoSb.

\begin{figure}[htb]
   \centering
   \includegraphics[width=8cm]{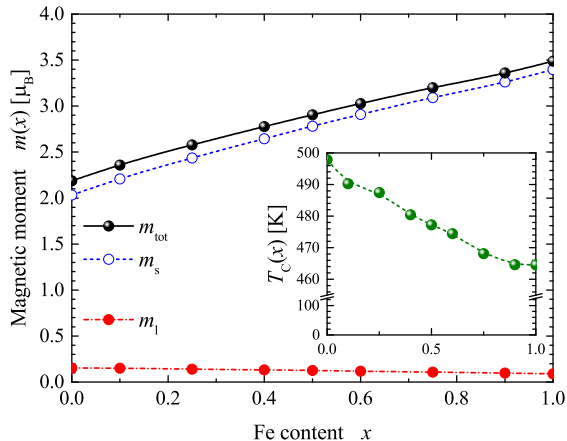}
   \caption{Magnetic properties of Rh$_2$Fe$_{x}$Co$_{1-x}$Sb.\\
            Total ($m_{\rm tot}$), spin ($m_s$), and orbital ($m_l$)
            magnetic moments as functions of Fe content $x$ are shown.
            The inset shows the Curie temperature ($T_C$). }
   \label{fig:mFeCo}
\end{figure}

The dependence of the magnetic properties on the composition is shown in
Figure~\ref{fig:mFeCo}. The total magnetic moment increases with increasing Fe
content, mainly because Fe has a higher spin magnetic moment
($\approx3\:\mu_B$) than Co ($\approx1.7\:\mu_B$). The individual
magnetic moments of the atoms are nearly unaffected by the composition. The
calculated Curie temperature decreases with increasing Fe content.

%%%%%%%%%%%%%%%%%%%%%%%%%%%%%%%%%
%%%%%%%%%%%%%%%%%%%%%%%%%%%%%%%%%%%%%%%

%%%%%%%%%%%%%%%%%%%%%%%%%%%%%%%%%%%%%%%%%%%%%%%%%%%%%%%%%%%%%%%%%%%%%%%%%%%%%%%
\subsection{Magnetocrystalline anisotropy of Rh$_2$Fe$_{x}$Co$_{1-x}$Sb}
\label{sec:mca}

Thus far, only the simplest case of uniaxial magnetocrystalline anisotropy has
been considered. The equations for extending the calculations to more detailed
cases are given in Appendix~\ref{app:app1}. These equations were used to
calculate the fourth-order uniaxial and tetragonal anisotropy constants, which
were used to obtain the magnetocrystalline anisotropy energy distributions.

\begin{figure}[htb]
   \centering
   \includegraphics[width=8cm]{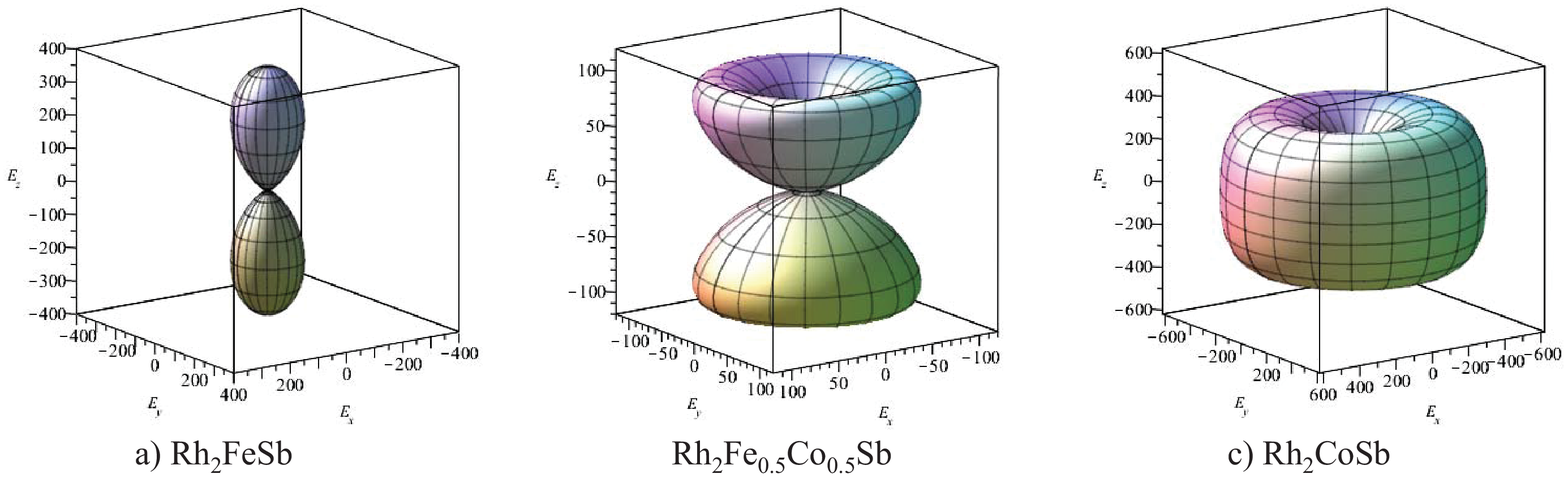}
   \caption{Uniaxial magnetic anisotropy of Rh$_2T$Sb compounds. \\
            Energy distributions $E_{u'}(\theta,\phi)$
            of $T=$~Fe (a), Fe$_{0.5}$Co$_{0.5}$~(b), and~Co~(c).
            The energies $E_{x,y,z}$ are given in $\mu$eV.
            (Please note the different energy scales.) }
   \label{fig:uniani}
\end{figure}

The calculated uniaxial energy distributions $E_{u'}(\theta,\phi)$ (see
Equations~(\ref{eq:uni}) and~(\ref{eq:eap}) in the Appendix) of Rh$_2$FeSb,
Rh$_2$Fe$_{0.5}$Co$_{0.5}$Sb, and Rh$_2$CoSb are plotted in
Figure~\ref{fig:uniani} for comparison.

%%%%%%%%%%%%%%%%%%%%%%%%%%%%%%%%%
%%%%%%%%%%%%%%%%%%%%%%%%%%%%%%%%%%%%%%%

The different behaviour of the anisotropy is clearly revealed in
Figure~\ref{fig:uniani}. Rh$_2$FeSb has an {\it easy} plane, and $c$ is the
{\it hard} axis; Rh$_2$Fe$_{0.5}$Co$_{0.5}$Sb has an {\it easy} plane as well, but
a {\it hard} cone, and in Rh$_2$CoSb, the $c$ direction is the {\it easy} axis.
Rh$_2$Fe$_{0.5}$Co$_{0.5}$Sb has a much lower anisotropy than the pure compounds, and
the differences between the energies of the $ab$ plane and the $c$ axis are very
small. A hard cone appears with its maximum at an angle of
$\theta_{3,4}=\pm35.7^\circ$ (see Equation~(\ref{eq:thetai}) in
Appendix~\ref{app:app1_uni}).

The calculated anisotropy constants for uniaxial and tetragonal symmetry are
compared in Table~\ref{tab:compKaniso}. The simple $K_u$ from
Equation~(\ref{eq:simple}) (see Appendix~\ref{app:app1}) clearly cannot describe
the magnetic anisotropy correctly.

% Table N %%%%%%%%%%%%%%%%%%%%%%%%%%%%%%%%%%%%%%%%%%%%%%%%%%%%%%%%%%%%%%%%%%%%%%%
\begin{table}[htb]
\centering
    \caption{Comparison of the anisotropy constants of Rh$_2T$Sb, $T=$~Fe, Fe$_{0.5}$Co$_{0.5}$, and Co.}
    \begin{ruledtabular}
    \begin{tabular}{l ccc}
                             & Rh$_2$FeSb & Rh$_2$Fe$_{0.5}$Co$_{0.5}$Sb & Rh$_2$CoSb \\
       \hline
       uniaxial \\
       $K_u$ [MJ/m$^3$]      & -1.21  &  -0.23 &  1.37      \\
       \hline
       $K_0$ [MJ/m$^3$]      &  1.31  &   0.39 &  0.0       \\
       $K_2$ [MJ/m$^3$]      & -2.19  &   0.50 &  3.62      \\
       $K_4$ [MJ/m$^3$]      &  0.98  &  -0.73 & -2.25      \\
       \hline
       tetragonal \\
       $K_{0,0}$ [MJ/m$^3$]  &  1.31  &   0.39 &  0.0       \\
       $K_{2,0}$ [MJ/m$^3$]  & -2.19  &   0.50 &  3.62      \\
       $K_{4,0}$ [MJ/m$^3$]  &  0.93  &  -0.81 & -2.40      \\
       $K_{4,4}$ [MJ/m$^3$]  &  0.05  &   0.08 &  0.15      \\
    \end{tabular}
    \end{ruledtabular}
    \label{tab:compKaniso}
\end{table}
%%%%%%%%%%%%%%%%%%%%%%%%%%%%%%%%%%%%%%%%%%%%%%%%%%%%%%%%%%%%%%%%%%%%%%%%%%%%%%%%%

\begin{figure}[htb]
   \centering
   \includegraphics[width=8cm]{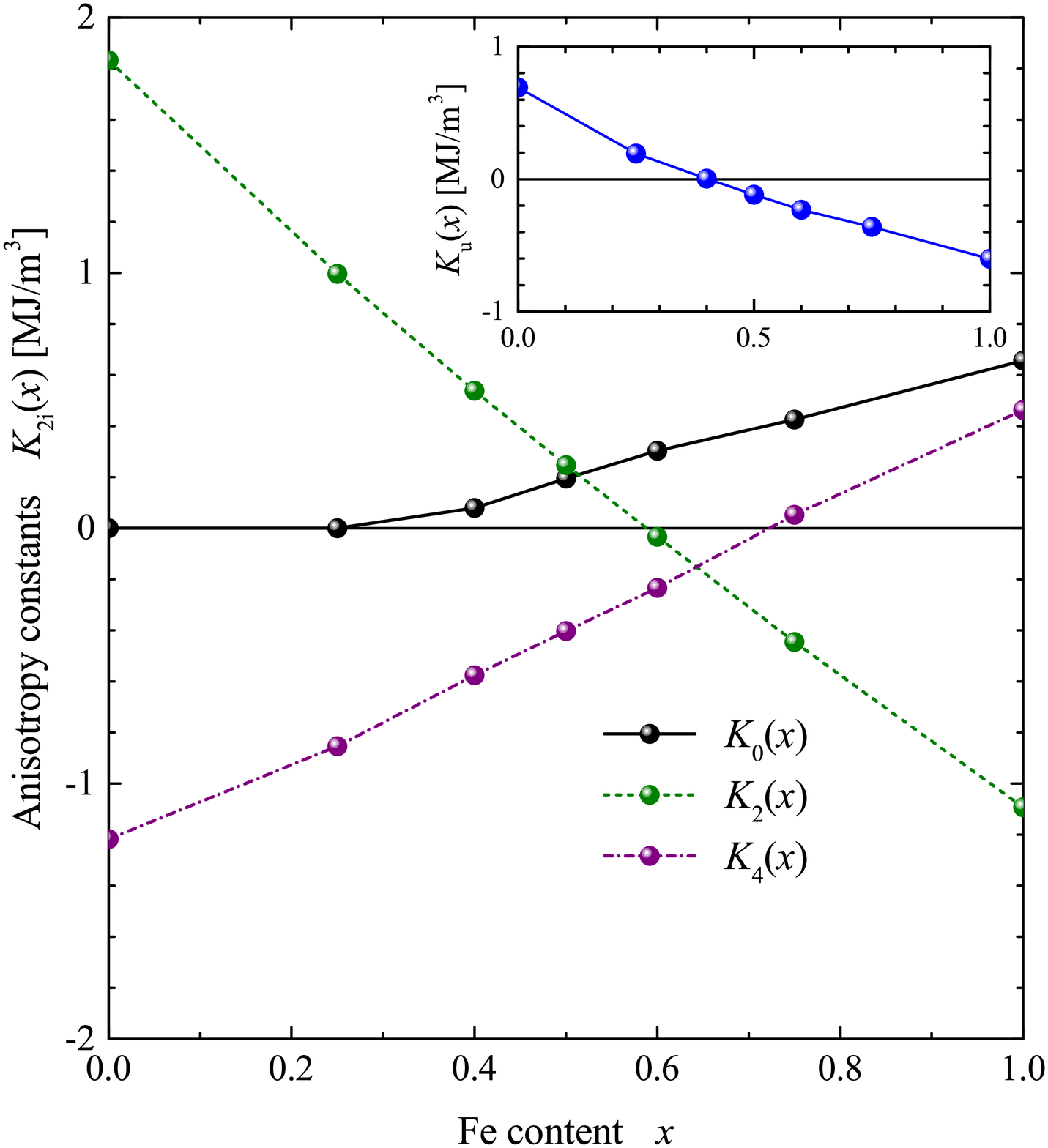}
   \caption{Anisotropy constants of Rh$_2$Fe$x$Co$_{1-x}$Sb compounds. \\
            The inset shows the uniaxial anisotropy constant obtained using
            Equation~(\ref{eq:simple2}) in Appendix~\ref{app:app1_uni}.}
   \label{fig:anicomposition}
\end{figure}

\begin{figure}[htb]
   \centering
   \includegraphics[width=8cm]{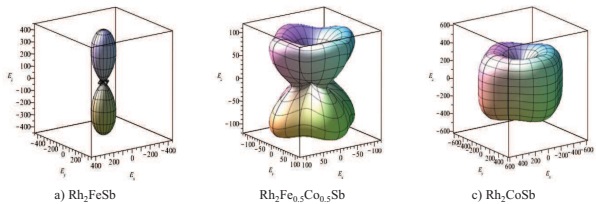}
   \caption{Tetragonal magnetic anisotropy of Rh$_2T$Sb compounds. \\
            Energy distributions $E_{a'}(\theta,\phi)$
            of $T=$~Fe (a), Fe$_{0.5}$Co$_{0.5}$~(b), and~Co~(c).
            The energies $E_{x,y,z}$ are given in $\mu$eV.
            (Please note the different energy scales.) }
   \label{fig:tetraani}
\end{figure}

The dependence of the uniaxial anisotropy constants on the composition is
illustrated in Figure~\ref{fig:anicomposition}. The uniaxial anisotropy constant
$K_u$ decreases with increasing iron content and exhibits a zero-crossing at
$x_0\approx0.4$. At intermediate iron contents, more complex behaviour appears,
as shown by the composition dependence of $K_{2i}$ and the results in
Figures~\ref{fig:uniani} and~\ref{fig:tetraani}.

%%%%%%%%%%%%%%%%%%%%%%%%%%%%%%%%%
%%%%%%%%%%%%%%%%%%%%%%%%%%%%%%%%%%%%%%%

The calculated tetragonal energy distributions $E_{a'}(\theta,\phi)$ (see
Equations~(\ref{eq:etetra}) and~(\ref{eq:eap}) in the Appendix) of Rh$_2$FeSb,
Rh$_2$Fe$_{0.5}$Co$_{0.5}$Sb, and Rh$_2$CoSb are shown in
Figure~\ref{fig:tetraani}. As in the plot of the uniaxial anisotropy in
Figure~\ref{fig:uniani}, the differences in the anisotropy are easily observed.
In Rh$_2$FeSb, the {\it hard} axis is along the $z$ ($[001]$) direction, and the
anisotropy exhibits weak variation in the basal plane, which is close to the
{\it easy} plane. Closer examination of the basal plane shows biaxial behaviour
with {\it easy} axes along the $[110]$ and $[1\overline{1}0]$ axes, but the
energy difference between these directions and the $[100]$ or $[010]$ axes is
very small. The anisotropy of Rh$_2$CoSb is still almost uniaxial, with the
{\it easy} axis along the $c$ ($[001]$) axis, and varies weakly in the basal
plane. Rh$_2$Fe$_{0.5}$Co$_{0.5}$Sb has much lower anisotropy than the pure
compounds and exhibits more complicated directional behaviour.

%%%%%%%%%%%%%%%%%%%%%%%%%%%%%%%%%
%%%%%%%%%%%%%%%%%%%%%%%%%%%%%%%%%%%%%%%

\begin{figure}[htb]
   \centering
   \includegraphics[width=8cm]{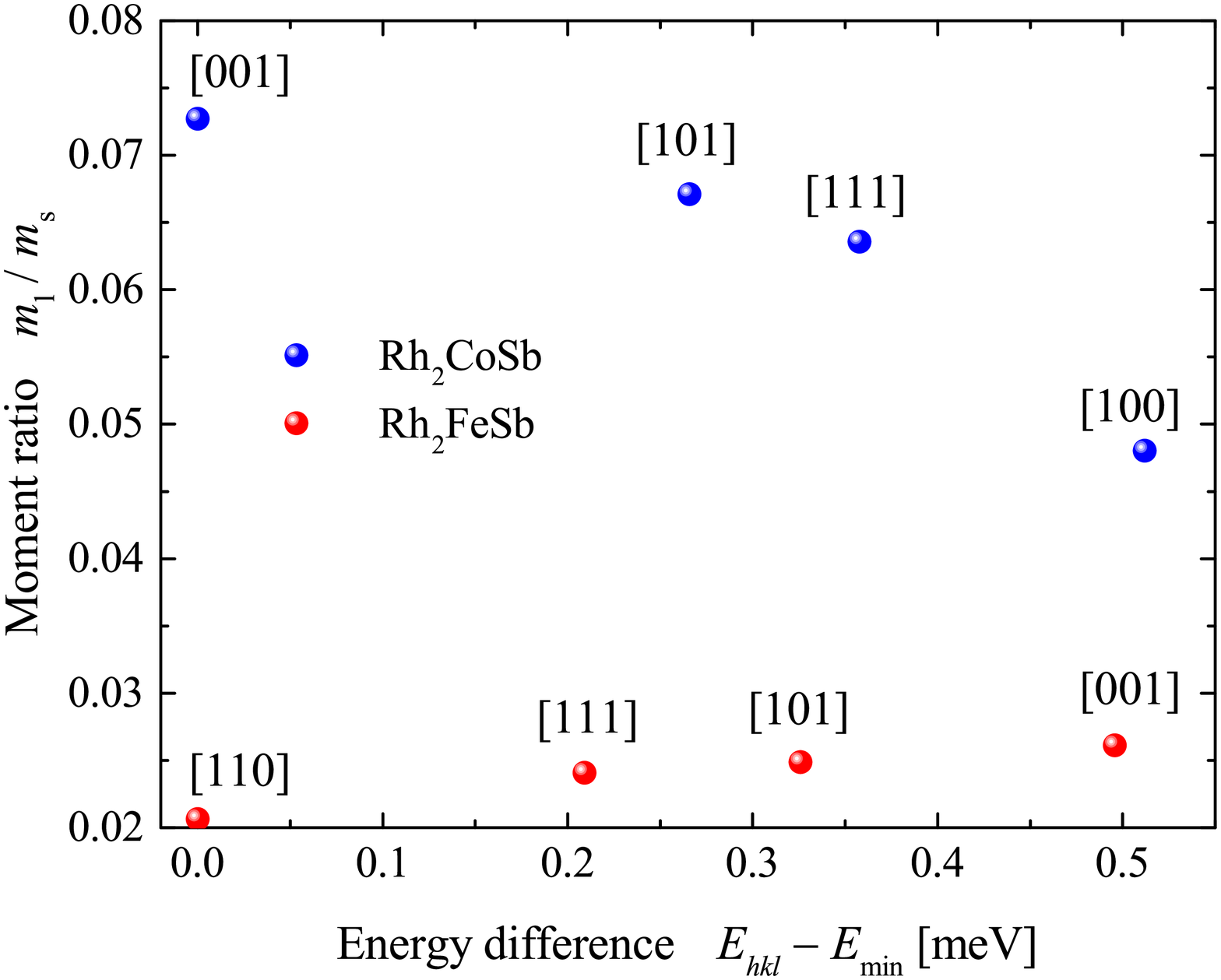}
   \caption{Directional dependence of the orbital moments of Rh$_2T'$Sb, $T'=$~Co, Fe.\\
            Note that the orbital moments are given relative
            to the spin moments for better comparison.}
   \label{fig:ml-ms}
\end{figure}

The directional dependence of the orbital magnetic moments was analysed to
clarify the role of the spin--orbit interaction. The magnetic moments for $m\|c$
are listed in Table~\ref{tab:compCoFe}. The ratio of the total orbital moment
to the total spin moment, $m_l/m_s$, was used owing to the large differences
between the magnetic moments for different compositions. Figure~\ref{fig:ml-ms}
shows the ratio $m_l/m_s$ as a function of the difference in the energies in
several magnetisation directions $[hkl]$. For both Rh$_2$CoSb and Rh$_2$FeSb,
the ratio is largest for magnetisation along the $c$ axis ($[001]$) and lowest
in the basal plane. This finding involves not only the ratio but also the
orbital momenta themselves, indicating that the orbital moment is not always
largest when the magnetisation is along the easy axis (or in the easy plane).
Here it depends at least partially on the angle between the magnetisation and
$c$ axis, as shown by the values for other directions.

%%%%%%%%%%%%%%%%%%%%%%%%%%%%%%%%%
%%%%%%%%%%%%%%%%%%%%%%%%%%%%%%%%%%%%%%%

\begin{figure*}[htb]
   \centering
   \includegraphics[width=16cm]{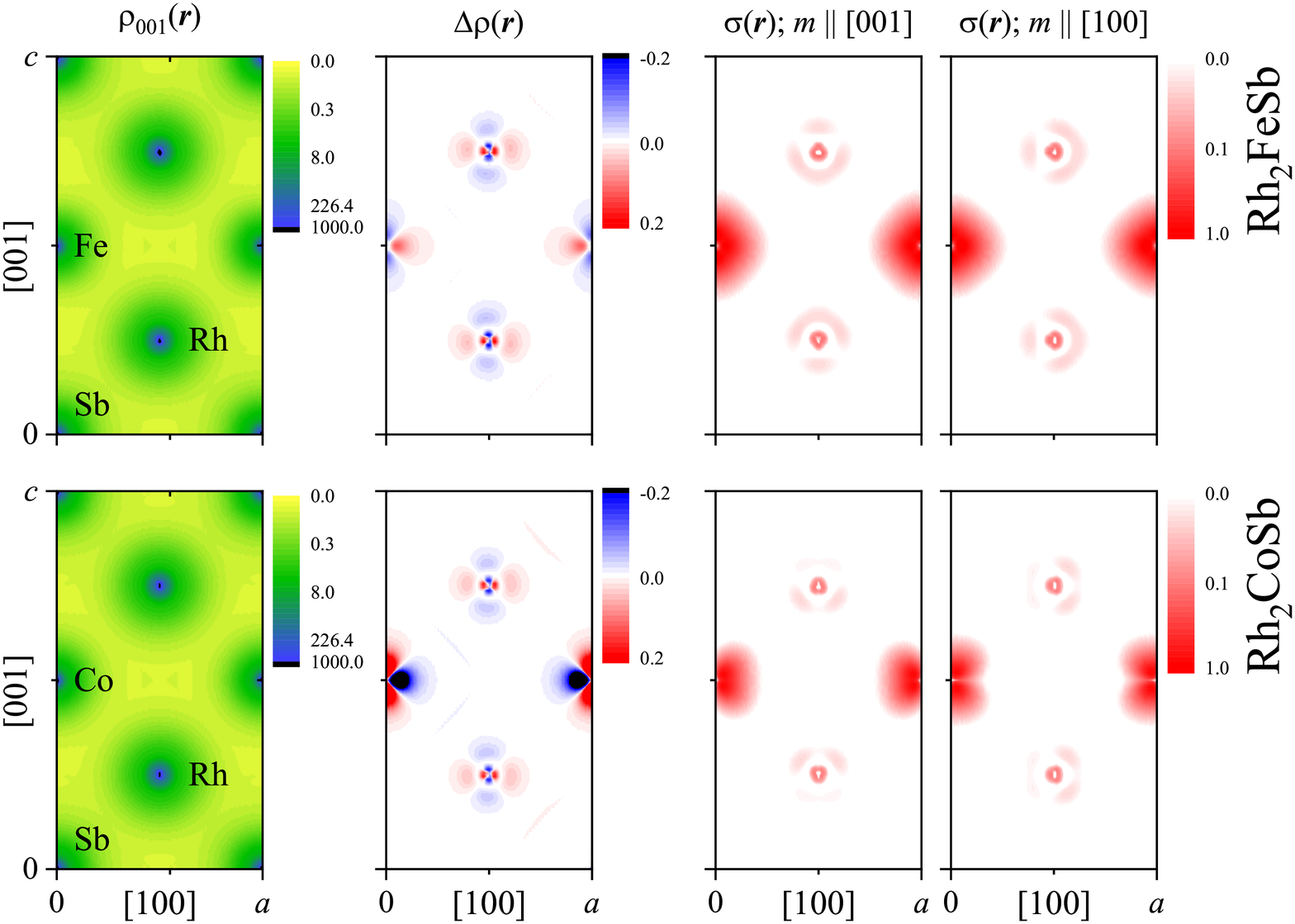}
   \caption{Electronic structure of Rh$_2$FeSb and Rh$_2$CoSb.\\
            Fully relativistic charge ($\rho(r)$) and spin ($\sigma(r)$) density distributions
            in a (100)-type plane for magnetisation parallel and perpendicular to the $c$ axis are shown.
            $\rho_{001}(r)$ is the charge density for $m\|[001]$, and $\Delta\rho=\rho_{001}-\rho_{100}$ is the difference
            between the charge densities for $m\|[001]$ and $m\|[100]$.
            (Note: colour bars are in atomic units.) }
   \label{fig:chargespin}
\end{figure*}

To further examine the nature of the anisotropy, the charge and spin density
distributions were analysed with respect to the magnetisation direction (compare
also Figures~\ref{fig:chargeCo} and~\ref{fig:chargeFe}). As mentioned above, the
symmetry changes when the magnetisation is applied along different
crystallographic directions. The point group symmetry of the 2b sites occupied
by Fe and Co is $D_{4h}$ and $D_{2d}$ for Rh on 4d. Applying the
magnetisation along one of the high-symmetry axes, i.e., the $c$ ([001]) or $a$
($[100]$) axis, changes the symmetry of the 2b sites to $C_{4h}$ or $C_{2h}$,
respectively. As a result, the irreducible representations and basic functions
depend on the magnetisation direction. For $C_{4h}$, they are $a_g$, $b_g$, and
$e_g$ with the $l=2$ basic functions $d_{z^2}$, ($d_{x^2-y^2}$, $d_{xy}$), and
($d_{xz}$, $d_{yz}$). For $C_{2h}$, they are $a_g$ and $b_g$ with ($d_{z^2}$,
$d_{x^2-y^2}$, $d_{xy}$) and ($d_{xz}$, $d_{yz}$). Similar differences appear
for the 4d sites. The charge and spin density distributions for different
magnetisation directions are compared in Figure~\ref{fig:chargespin} for the
compounds containing only Fe or Co.

%%%%%%%%%%%%%%%%%%%%%%%%%%%%%%%%%
%%%%%%%%%%%%%%%%%%%%%%%%%%%%%%%%%%%%%%%

As mentioned above, the details of the charge density are not easily observed
directly from the graph when the magnetisation direction is changed, because the
graph shows mainly the positions of the atoms. However, they can be observed if
one investigates the difference in the charge distribution, which is plotted as
$\Delta\rho(r)$. It was calculated for both compounds as the difference between
the charge densities obtained assuming that the magnetisation is parallel
($m\|[001]$) or perpendicular ($m\|[100]$) to the $c$ axis. In both compounds,
the magnetisation has the same effect on $\Delta\rho(r)$ near the Rh atoms. That
is, the charge distribution is rotated with the direction of the magnetisation.
In the same way, the Rh-based spin densities are affected by the magnetisation
direction. They change from [001]-aligned when $m\|[001]$ to [100]-aligned when
$m\|[100]$, regardless of which $3d$ transition metal is used. The situation is
different near the $3d$ transition metals Fe and Co, where $\Delta\rho(r)$ and
$\sigma(r)$ are affected very differently by the magnetisation direction. The
reason is the different occupation of $3d$ valence electrons of Fe
($n_d^{\rm Fe}=6.6$) and Co ($n_d^{\rm Co}=7.8$), which are responsible for the
different spin moments. The overall differences in the charge and spin densities
at different magnetisation directions result in different total energies.

Finally, the gain or loss of energy with changes in the magnetisation direction
results in the magnetocrystalline anisotropy. The electronic structure of the
two compounds, Rh$_2$FeSb and Rh$_2$CoSb, differs depending on the magnetisation
direction, which is reflected in the change in the anisotropy from the easy
plane to the easy axis when Fe is replaced with Co.

%%%%%%%%%%%%%%%%%%%%%%%%%%%%%%%%%%%%%%%%%%%%%%%%%%%%%%%%%%%%%%%%%%%%%%%%%%%%%%%
\section{Conclusions}

The electronic and magnetic structure of tetragonal Heusler
compounds with the composition Rh$_2$Fe$_{x}$Co$_{1-x}$Sb were investigated by
{\it ab initio} calculations. The calculations revealed that the magnetic
moment increases and the Curie temperature decreases with increasing Fe content
$x$. The Rh atoms have only small, composition-independent magnetic moments.
The magnetic properties are determined by those of the Fe and Co atoms and thus
depend strongly on the composition. The total energies for various magnetisation
directions were calculated to determine the magnetic anisotropy. The analysis is
described in detail in an extended Appendix. For bulk materials, the
magnetocrystalline anisotropy is found to be much stronger (by three orders of
magnitude) than the dipolar anisotropy. Special attention was given to the
borderline compounds, Rh$_2$FeSb and Rh$_2$CoSb. The most striking result was
that a composition-dependent transition from easy-axis to easy-plane anisotropy
occurs at an iron concentration of approximately 40\%.

%%%%%%%%%%%%%%%%%%%%%%%%%%%%%%%%%%%%%%%%%%%%%%%%%%%%%%%%%%%%%%%%%%%%%%%%%%%%%%%
\appendix

\section{Disorder}
\subsection{Rh$_2$CoSb with Co--Sb-type antisite disorder}
\label{app:app0}

Supplementary calculations were performed for disordered Rh$_2$CoSb and
Rh$_2$FeSb using {\sc Sprkkr} using the coherent potential approximation. For
example, the disordered compound may be written as
Rh$_2$(Co$_{1-x/2}$Sb$_{x/2}$)(Co$_{x/2}$Sb$_{1-x/2}$), where $x$ is the
disorder level. The result for $x=1$, which denotes complete Co--Sb disorder, is
illustrated in Figure~\ref{fig:struct}(b). Alternatively, it can be assumed that
disorder between the Co and Rh atoms decreases the magnetic moments, which is
consistent with the results of calculations of the inverted structure in space
group 119, but not with those when Co--Sb disorder is assumed, as shown below.

\begin{figure}[htb]
   \centering
   \includegraphics[width=8cm]{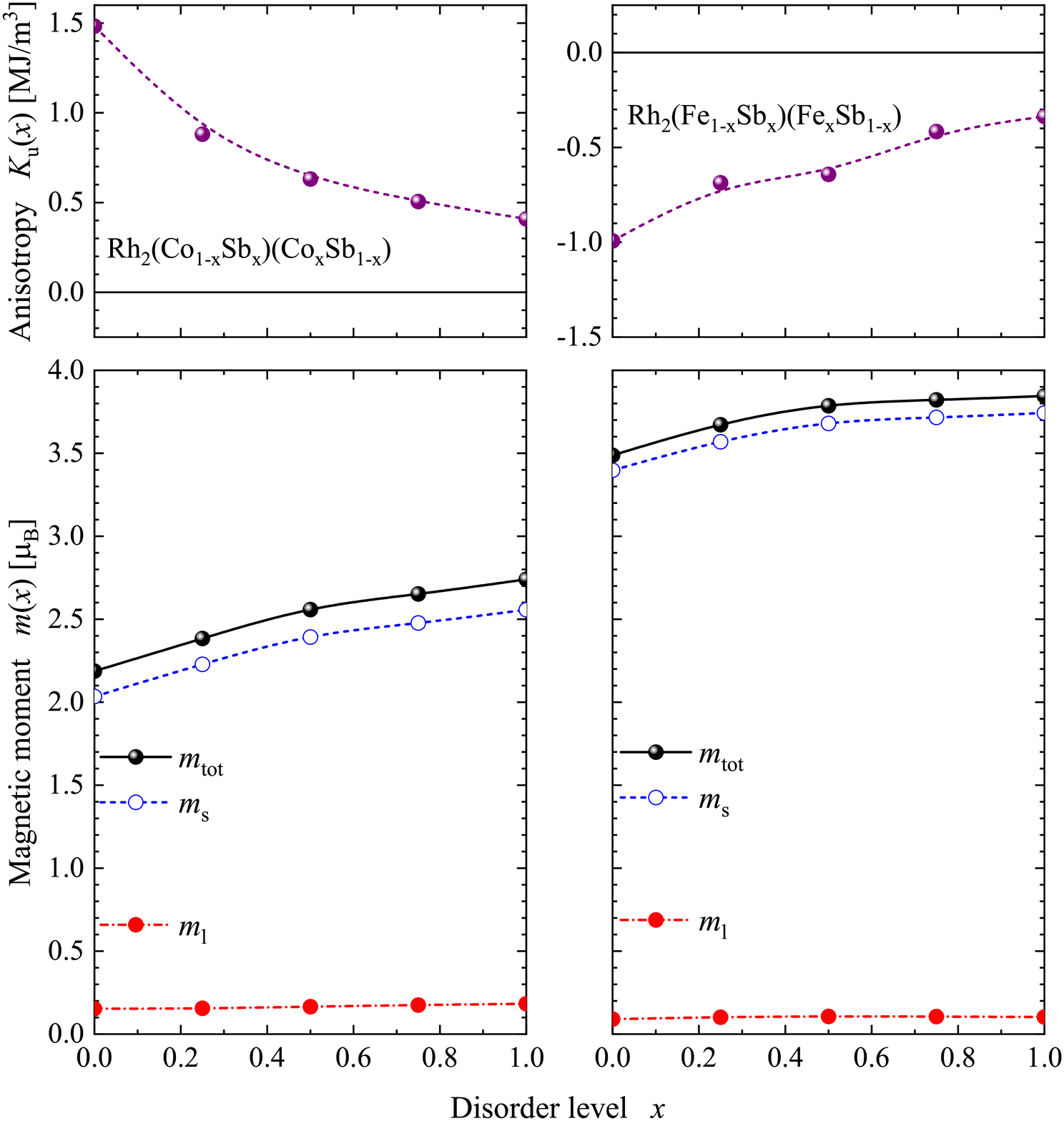}
   \caption{Disorder-induced changes in magnetic moments and anisotropy of
            Rh$_2$($T_{1-x/2}$Sb$_{x/2}$)($T_{x/2}$Sb$_{1-x/2}$) $T=$~Co, Fe. \\
            Total, spin, and orbital
            magnetic moments are shown, along with the uniaxial anisotropy
            constant $K_u$ as a function of disorder level $x$. }
   \label{fig:magmomdisorder}
\end{figure}

The evolution of the magnetic moments of Rh$_2$CoSb and Rh$_2$FeSb with
increasing disorder is shown in Figure~\ref{fig:magmomdisorder}. The total
magnetic moment in the fully disordered state is approximately 20\% larger for
Rh$_2$CoSb and approximately 10\% larger for Rh$_2$FeSb than those of the
compounds in the completely ordered state. The orbital moments are nearly
constant in both compounds and are independent of the degree of disorder ($x$).
The decrease in the total moments with decreasing $x$ is attributed to the
decrease in the spin magnetic moments of both compounds.

%%%%%%%%%%%%%%%%%%%%%%%%%%%%%%%%%
%%%%%%%%%%%%%%%%%%%%%%%%%%%%%%%%%%%%%%%

Figure~\ref{fig:magmomdisorder} shows that the type of anisotropy (easy axis for
Rh$_2$CoSb and easy plane for Rh$_2$FeSb) is retained even in the completely
disordered state. However, the absolute value of the second-order uniaxial
anisotropy constant $K_u$ decreases. That is, the anisotropy becomes weaker with
increasing disorder. The $K_u$ values of both compounds are approximately 70\%
higher in the completely disordered state. No direct correspondence is observed
between the behaviour of $K_u$ and that of the spin, orbital, or total magnetic
moments. The effects of disorder and composition on the magnetic anisotropy of
the Co--Fe system were investigated using first principles CPA calculations by
Turek {\it et al.}~\cite{TKC12}, who also observed a decrease in anisotropy with
increasing disorder.

%%%%%%%%%%%%%%%%%%%%%%%%%%%%%%%%%%%%%%%%%%%%%%%%%%%%%%%%%%%%%%%%%%%%%%%%%%%%%%%
\subsection{Off-stoichiometric alloys}
\label{sec:offst}

In many experiments, 2:1:1 stoichiometry was not fully reached, but an excess of
Fe or Co and a deficiency of Sb was obtained. In particular, the magnetic
properties of Rh$_2T_{1+x}$Sb$_{1-x}$, with $T=$~Fe, Co, were calculated for
$x=0.12$. As in the study of disorder, the calculations were performed using the
CPA. The $2a$ site is assumed to be occupied by 12\% with Fe (or Co) and by 88\%
with Sb, whereas the occupations of the $4d$ and $2b$ sites are unchanged.

The calculated magnetic properties of the off-stoichiometric alloys are listed
in Table~\ref{tab:compdisorder}. The magnetic moments and spin stiffness $D_0$
are enhanced in both alloys, and the values are higher than those of the
stoichiometric compound. In particular, the excess Co and Fe atoms on the $2a$
site contribute a large spin moment. The total magnetic moment,
$m_s+m_l=2.627\:\mu_B$, of Rh$_2$Co$_{1.12}$Sb$_{0.88}$ is very similar to the
experimentally observed value of $2.6\:\mu_B$. The Curie temperature of the off-
stoichiometric Co-containing compound is slightly lower, whereas that of the Fe
compound is slightly higher. These findings, along with the spin stiffness
results, suggest that the exchange coupling parameters of the stoichiometric
compounds differ from those of the off-stoichiometric alloys.

The type of anisotropy (easy plane or easy axis) is the same in the
off-stoichiometric alloys as in the stoichiometric compounds. The
uniaxial anisotropy constants differ, however. They are enhanced
in the Fe alloy and reduced in the Co alloy.

% Table N %%%%%%%%%%%%%%%%%%%%%%%%%%%%%%%%%%%%%%%%%%%%%%%%%%%%%%%%%%%%%%%%%%%%%%%
\begin{table}[htb]
\centering
    \caption{Calculated magnetic properties of off-stoichiometric Rh$_2T_{1.12}$Sb$_{0.88}$. \\
             Spin $m_s$ and orbital $m_l$ magnetic moments per atom (Rh, Co, Fe) and
             those of the primitive cell ($total$) are listed, as well as the Curie temperature $T_C$ and spin stiffness $D_0$.
             $m^{\rm 2b}_{s,l}$ represents the magnetic moments at the original position, and $m^{\rm 2a}_{s,l}$ represents
             the moments of the excess Fe and Co atoms at the initial Sb position. }
    \begin{ruledtabular}
    \begin{tabular}{l cc}
       Rh$_2T_{1.12}$Sb$_{0.88}$  & $T=$~Fe & $T=$~Co \\
       \hline
       $m_s^{\rm Rh}$ [$\mu_B$]   & 0.308   &  0.244  \\
       $m_l^{\rm Rh}$ [$\mu_B$]   & 0.012   &  0.009  \\
       $m_s^{\rm 2b}$ [$\mu_B$]   & 2.988   &  1.691  \\
       $m_l^{\rm 2b}$ [$\mu_B$]   & 0.092   &  0.140  \\
       $m_s^{\rm 2a}$ [$\mu_B$]   & 3.564   &  2.521  \\
       $m_l^{\rm 2a}$ [$\mu_B$]   & 0.072   &  0.155  \\
      \hline
       $m_s^{total}$ [$\mu_B$]    & 4.010   &  2.453  \\
       $m_l^{total}$ [$\mu_B$]    & 0.124   &  0.174  \\
      \hline
       $T_C$ [K]                  & 490     &   480   \\
       $D_0$ [meV~\AA$^2$]        & 690     &  1100   \\
       \hline
 %      $K_u$ [MJ/m$^3$]           & -2.22   & 0.82  \\  % fullpot sphercell check Co !!!
 %      $\mu_0 H_a$ [T]            & -11.08  & -3.8  \\  % fullpot sphercell check Co !!!
       $K_u$ [MJ/m$^3$]           & -1.667  & 0.826  \\  % fullpot sphercell
 %      $\mu_0 H_a$ [T]            &  3.698  & -3.803   \\  % fullpot sphercell
    \end{tabular}
    \end{ruledtabular}
    \label{tab:compdisorder}
\end{table}
%%%%%%%%%%%%%%%%%%%%%%%%%%%%%%%%%%%%%%%%%%%%%%%%%%%%%%%%%%%%%%%%%%%%%%%%%%%%%%%%%

%%%%%%%%%%%%%%%%%%%%%%%%%%%%%%%%%%%%%%%%%%%%%%%%%%%%%%%%%%%%%%%%%%%%%%%%%%%%%%%
\section{Semi-relativistic band structures}
\label{app:srbs}

\begin{figure}[htb]
   \centering
   \includegraphics[height=8cm]{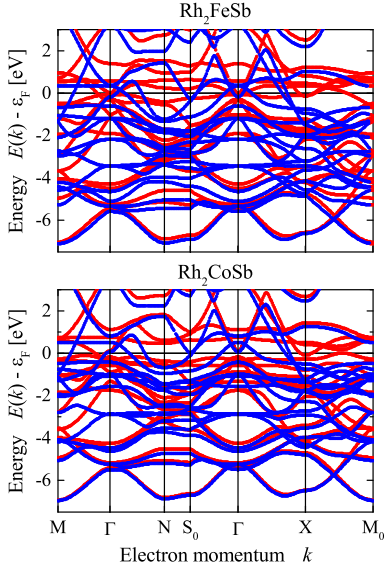}
   \caption{Semi-relativistic band structures of Rh$_2$FeSb and Rh$_2$CoSb. \\
            Red and blue indicate majority and minority states, respectively. }
   \label{fig:band-spsrel}
\end{figure}

The semi-relativistic band structures of Rh$_2$FeSb and Rh$_2$CoSb are compared
in Figure~\ref{fig:band-spsrel} to illustrate the spin characteristics of the
bands. The band structures are similar; the main differences result from the
larger band filling in the Co-based compound, which has one more valence
electron than the Fe compound. Further, the larger spin splitting in the Fe
compound clearly results in a large spin magnetic moment.

%%%%%%%%%%%%%%%%%%%%%%%%%%%%%%%%%
%%%%%%%%%%%%%%%%%%%%%%%%%%%%%%%%%%%%%%%

%%%%%%%%%%%%%%%%%%%%%%%%%%%%%%%%%%%%%%%%%%%%%%%%%%%%%%%%%%%%%%%%%%%%%%%%%%%%%%%
\section{Magnetocrystalline anisotropy}
\label{app:app1}

In this Appendix, the discussion of the magnetocrystalline anisotropy is
extended beyond simple uniaxial approximations. The magnetocrystalline energy of
uniaxial systems can be derived from the first principles total energies for
different magnetisation directions (quantisation axes).

Textbooks give different descriptions of the magnetic anisotropy, in particular,
different equations for the anisotropy
constants~\cite{SCo99,Kub00,CGr09,Coe10,Spa11,Kri16}. Therefore, care must be
taken when comparing the results of this work with those of other studies or
comparing other studies with each other.

% Kub00: cos(6 phi)
% CGr09,
% Spa11: nur cubic, uniaxial
% Kri16: alles, hex: sin(6 phi), tetra: sin^2(phi)*cos^2(phi)
% SCo99: hex: cos(6 phi), tetra: cos(4 phi)
% Coe10: alles, hex: sin(6 phi), tetra: cos(4 phi)

%%%%%%%%%%%%%%%%%%%%%%%%%%%%%%%%%%%%%%%%%%%%%%%%%%%%%%%%%%%%%%%%%%%%%%%%%%%%%%%
\subsection{Uniaxial magnetic anisotropy}
\label{app:app1_uni}

It is often assumed that the magnetocrystalline anisotropy in tetragonal or
hexagonal systems is simply described by a second-order dependence on the angle
$\theta$ between the $c$ axis and the magnetisation direction, that is,

\begin{equation}
\label{eq:secndord}
   K_u \sin^2(\theta),
\end{equation}

where $K_u$ is the uniaxial anisotropy constant. In that case,

\begin{equation}
\label{eq:simple}
  K_u = E^{100}-E^{001}
\end{equation}

is simply calculated from the difference between the energies for magnetisation
along the principal axes, $c\|[001]$ and $a\|[100]$. For $K_u>0$, the
{\it easy} axis is along the $c$ axis, whereas $K_u<0$ describes an {\it easy} plane where $c$
is the {\it hard} axis. For a distinct magnetic anisotropy in the
$ab$ plane, it would be more accurate to use the lowest energy of the two
in-plane directions along the principal axis and the diagonal, which are the [100]
and [110] directions, respectively:

\begin{equation}
\label{eq:simple2}
  K_u = \min(E^{100},E^{110})-E^{001}.
\end{equation}

Equation~(\ref{eq:secndord}) has another serious drawback; namely, the
anisotropy is completely independent of the crystal lattice, and the anisotropic
energy distribution always has the same shape regardless of the $c/a$ parameter
and whether the crystal has tetragonal, hexagonal, or some other structure. That
is, Equation~(\ref{eq:secndord}) is ultimately useful only for distinguishing
between {\it easy} and {\it hard} $c$ axes.

Now we consider only tetragonal systems. By using the series expansion
$\sum K_{2\nu,0} \sin^{2\nu}(\theta)$ up to the fourth order in
$\sin(\theta)$, the uniaxial magnetocrystalline energy is expressed as

\begin{equation}
\label{eq:uni}
  E^{\rm uniaxial}_{\rm crys} = K_0 + K_2 \sin^{2}(\theta) + K_4 \sin^4(\theta).
\end{equation}

The equations for the sixth-order uniaxial anisotropy are discussed by Jensen
and Bennemann~\cite{JBe06}, for example. In the following, the subscript "crys"
is omitted, and the energies are indexed only by direction or by "uni".  For the
high-symmetry directions [$h,k,l$] and the lowest indices ($h,k,l=0,1$), the
energies depend on the anisotropy coefficients as follows:

\begin{eqnarray}
\label{eq:hkl}
 E^{001} & = & K_0,  \\
 E^{100} & = & K_0 + K_2 + K_4, {\rm or}   \nonumber \\
 E^{110} & = & K_0 + K_2 + K_4, {\rm and}  \nonumber \\
 E^{101} & = & K_0 + K_2 \sin^{2}(\theta^{101}) + K_4 \sin^{4}(\theta^{101}), {\rm or} \nonumber \\
 E^{111} & = & K_0 + K_2 \sin^{2}(\theta^{111}) + K_4 \sin^{4}(\theta^{111}).          \nonumber
\end{eqnarray}

Note that the energies for the [100] and [110] directions are identical only
when uniaxial anisotropy is assumed. The energies for the [101] and [111]
directions, however, have different angles with respect to the
$c$ axis. From Equations~(\ref{eq:uni} and~\ref{eq:hkl}), $K_2$ and $K_4$ may be
obtained, for example, from the differences:

\begin{eqnarray}
\label{eq:diff}
 E^{100} - E^{001} & = & K_2 + K_4 \: {\rm and}  \\
 E^{101} - E^{001} & = & K_2 \sin^{2}(\theta) + K_4 \sin^{4}(\theta). \nonumber
\end{eqnarray}

For  $z=c/a$, the angle $\theta$ is found using
$\theta^{101} = \theta^{011} = \arctan(1/z)$. From Equation~(\ref{eq:hkl})
or~(\ref{eq:diff}), the anisotropy constants $K_i$ are given by

\begin{eqnarray}
\label{eq:ki}
 K_0 & = & E^{001}, \\
 K_2 & = & (E^{101} - E^{001})(z^2+2) + (E^{101} - E^{100})\frac{1}{z^2},    \nonumber \\
 K_4 & = & (E^{001} - E^{101})(z^2+1) + (E^{100} - E^{101})\frac{ (z^2 + 1)}{z^2}. \nonumber
\end{eqnarray}

Alternatively, $E^{111}$ and $\theta^{111} = \arctan(\sqrt{2}/z)$ may be used,
but the resulting equations will have a different dependence on $c/a$. The
uniaxial magnetocrystalline anisotropy energy ($E_{u}$) is the difference
between the magnetocrystalline energy (here $E^{\rm uni}$) and the isotropic
contribution, which is the spherical part $K_0$:

\begin{equation}
  E_{u} = E^{\rm uni} - K_0.
\label{eq:eu}
\end{equation}

According to this equation, the uniaxial magnetocrystalline anisotropy energy may be
positive or negative, depending on the directions and values of $K$
(see also Appendix~\ref{app:app1_plot}).

Equation~(\ref{eq:uni}) has four extremal values at

\begin{equation}
   \theta_i = 0, \frac{\pi}{2}, {\rm and}  \pm \arcsin \left( \sqrt{\frac{-K_2}{2 K_4}} \right),
\label{eq:thetai}
\end{equation}

where the first derivative of the fourth-order equation
[Equation~(\ref{eq:uni})] vanishes, that is, for $dE^{\rm uniaxial}/d\theta=0$.
The solutions $\theta_{3,4}$ are real only if the anisotropy constants obey the
relation $0\leq \frac{-K_2}{2 K_4} \leq1$, that is, $K_2K_4\geq0$,
$|K_2|\leq2|K_4|$. For $K_2=-2K_4$, one has $\theta_{3,4}=\pm90^\circ$. For a
real $\theta_{3,4}$, one has an {\it easy} or a {\it hard} cone. The resulting
extremal energies are

\begin{eqnarray}
\label{eq:eminmax}
 E(0)             & = & K_0,                       \\
 E(\pi/2)         & = & K_0 + K_2 + K_4, \nonumber      \\
 E(\theta_{3,4})  & = & K_0 - \frac{K_2^2}{4K_4}. \nonumber
\end{eqnarray}

The minima or maxima are obtained using the second derivatives of the energy at
the extremal angles:

\begin{eqnarray}
\label{eq:d2edt2}
 \left. \frac{d^2E}{d\theta^2} \right|_{0}            & = & 2 K_2,                       \\
 \left. \frac{d^2E}{d\theta^2} \right|_{\pi/2}        & = & -2 (K_2+2K_4), \nonumber      \\
 \left. \frac{d^2E}{d\theta^2} \right|_{\theta_{3,4}} & = & -\frac{2K_2(K_2+2K_4)}{K_4}. \nonumber
\end{eqnarray}

The minima appear for positive $2^{\rm nd}$ derivatives
($\left. d^2E/d\theta^2 \right|_{\theta_{i}}>0$) and define the easy
direction(s) of magnetisation. Indeed, one has to search for the absolute
minimum and maximum to find the correct easy and hard axes, planes, or cones. An
easy cone appears for $K_2<0$, $K_4>- K_2/2$, and the corresponding cone angle
is given by $\theta_{3,4}$. A special hard cone exists for $K_2>0$, $K_4=-K_2$,
where both the $c$ axis and the $ab$ plane have the same (lowest) energy. The
energy barrier at the hard cone must be overcome, however, to change the
magnetisation direction from the easy axis to the easy plane and vice versa. In
the range $-\infty<K_4<-K_2/2$, the solutions are metastable when $K_2>0$. The
complete fourth-order uniaxial anisotropy phase diagram is presented in
Table~\ref{tab:uniphasedia}.

% Table N %%%%%%%%%%%%%%%%%%%%%%%%%%%%%%%%%%%%%%%%%%%%%%%%%%%%%%%%%%%%%%%%%%%%%%%
\begin{table}[htb]
\centering
    \caption{Uniaxial anisotropy phase diagram. \\
             $ab$ stands for basal plane, $c$ stands for $c$-axis.
            Cones may have an opening angle $\theta$ or $\pi/4$ with respect to the $c$ axis. }
    \begin{ruledtabular}
    \begin{tabular}{l ccc}
      $K_2$   & $K_4$                   &     easy              &      hard      \\
      \hline
      $>0$    & $-\infty \cdots -K_2$   &     $ab$              &      cone ($\theta$)   \\
      $>0$    & $-K_2$                  &     $ab, c$           &      cone ($45^\circ$) \\
      $>0$    & $-K_2 \cdots -K_2/2$    &     $c$               &      cone ($\theta$)   \\
      $>0$    & $-K_2/2 \cdots \infty$  &     $c$               &      $ab$      \\
      \hline
      $=0$    & $=K_2=0$                & \multicolumn{2}{c}{undefined, spherical} \\
      \hline
      $<0$    & $-K_2 \cdots \infty$    &     cone ($\theta$)   &      $ab$      \\
      $<0$    & $-K_2$                  &     cone ($45^\circ$) &      $ab, c$   \\
      $<0$    & $-K_2 /2 \cdots -K_2$   &     cone ($\theta$)   &      $c$       \\
      $<0$    & $-\infty \cdots -K_2/2$ &     $ab$              &      $c$       \\
    \end{tabular}
    \end{ruledtabular}
    \label{tab:uniphasedia}
\end{table}
%%%%%%%%%%%%%%%%%%%%%%%%%%%%%%%%%%%%%%%%%%%%%%%%%%%%%%%%%%%%%%%%%%%%%%%%%%%%%%%%%

\begin{figure}[htb]
   \centering
   \includegraphics[height=8cm]{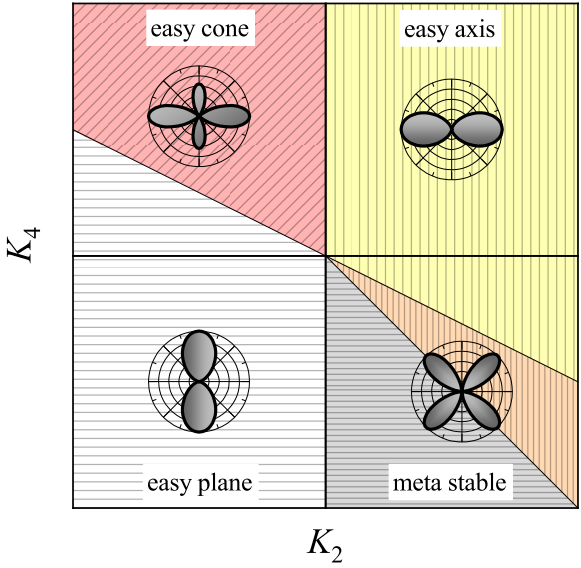}
   \caption{Magnetic anisotropy phase diagram. \\
            In the {\it metastability} range, {\it hard-cone}-type anisotropy occurs.
            In the sketches of the $E(\theta)$ distributions, it is assumed that $K_0=0$.}
   \label{fig:phasediag}
\end{figure}

The magnetic anisotropy phase diagram for fourth-order uniaxial anisotropy is
displayed in Figure~\ref{fig:phasediag}. It is similar to the graphical
representations reported in References~\cite{SOK98,JBe06}. The different phases
are distinguished. For $K_4=-K_2<0$, there is a distinct metastable case with
equal energies for magnetisation along the $c$ axis and in the $ab$ plane. At
this line, a transition occurs from easy-axis to easy-plane behaviour. In the
metastable region for $K_4<-K_2/2<0$, easy-axis behaviour appears, whereas easy-
plane behaviour appears for $K_4<-K_2<0$ (see Table~\ref{tab:uniphasedia}). In
both cases, the sizes of the anisotropy constants determine how easily one state
can switch to the other and the stability of the state with lower energy. The
energy barrier to cross the hard cone has a size of $-\frac{K_2^2}{4K_4}$, as
mentioned above.

%%%%%%%%%%%%%%%%%%%%%%%%%%%%%%%%%
%%%%%%%%%%%%%%%%%%%%%%%%%%%%%%%%%%%%%%%

%%%%%%%%%%%%%%%%%%%%%%%%%%%%%%%%%%%%%%%%%%%%%%%%%%%%%%%%%%%%%%%%%%%%%%%%%%%%%%%
\subsection{Tetragonal magnetic anisotropy}
\label{app:app1_tetra}

The uniaxial magnetic anisotropy does not reflect the symmetry of the crystal
structure. The symmetry of the anisotropy should generally be the same as the
symmetry of the crystal potential; thus, it is given by the fully symmetric
irreducible representation of the point group, that is, $a$, $a_1$, $a_g$, or
similar. Again, by using a series expansion up to the fourth order in
$\sin(\theta)$, the magnetocrystalline energy of a tetragonal system is
expressed as

\begin{eqnarray}
  E^{\rm tetragonal}_{\rm crys} & = & \sum_{\nu=0}^{2} K_{2\nu,0} \sin^{2\nu}(\theta) + K_{4,4} \sin^4(\theta)f(\phi), \nonumber \\
                                & = & E^{\rm uniaxial}_{\rm crys} + K_{4,4} \sin^4(\theta)f(\phi), \nonumber \\
                        f(\phi) & = & \cos(4\phi). \nonumber
\label{eq:etetra}
\end{eqnarray}

$f(\phi)$ has an azimuthal dependence on $4\phi$, which results in the expected
fourfold symmetry. Some works used $f'(\phi)=\sin^4(\phi)+\cos^4(\phi)$,
which results in different equations and $K$ values. Higher-order
approximations will include terms with $K_{6,0}$, $K_{6,4}$, $K_{8,0}$,
$K_{8,4}$, $K_{8,8}$, and so on. Subtracting $K_0$ from Equation~(\ref{eq:etetra}) yields

\begin{equation}
  E_a(\vec{r}) = K_{2,0} \sin^{2}(\theta) + [K_{4,0} + K_{4,4}\cos(4\phi)]\sin^{4}(\theta). %\nonumber
\label{eq:eetetra}
\end{equation}

In the following, the subscript "crys" is omitted, and the energies are
indexed only by direction or by "tet". For the high-symmetry directions
[$h,k,l$] and the lowest indices
($h,k,l=0,1$), Equation~(\ref{eq:etetra}) gives

\begin{eqnarray}
\label{eq:etethkl}
 E^{001} & = & K_{0,0}, \\
 E^{100} & = & K_{0,0} + K_{2,0} + K_{4,0} + K_{4,4}, \nonumber  \\
 E^{110} & = & K_{0,0} + K_{2,0} + K_{4,0} - K_{4,4},  {\rm and}  \nonumber  \\
 E^{101} & = & \sum_{{\nu=0,2}} K_{2\nu,0} \sin^{2\nu}(\theta^{101}) + K_{4,4} \sin^{4}(\theta^{101}), {\rm or} \nonumber  \\
 E^{111} & = & \sum_{{\nu=0,2}} K_{2\nu,0} \sin^{2\nu}(\theta^{111}) - K_{4,4} \sin^{4}(\theta^{111}).          \nonumber
\end{eqnarray}

For $z=c/a$, the angle $\theta^{101}$ is found using
$\theta^{101} = \theta^{011} = \arctan(1/z)$. Alternatively, $E^{111}$ with
$\theta^{111} = \arctan(\sqrt{2}/z)$ may be used. From the first four energies
of Equation~(\ref{eq:etethkl}), the anisotropy constants $K_{l,m}$ are found to be

\begin{eqnarray}
\label{eq:khkl}
 K_{0,0} & = & E^{001},  \\
%%%% for \cos(4 \phi)
 K_{2,0} & = & (E^{101} - E^{001}) (z^2 + 2)      \nonumber  \\
         &   & +(E^{101} - E^{100})\frac{1}{z^2}, \nonumber  \\
 K_{4,0} & = & (E^{001} - E^{101})(z^2 + 1)          \nonumber  \\
         &   & + (E^{100} - E^{101}) \frac{1}{z^2}        \nonumber  \\
         &   & + \frac{1}{2} (E^{100} +E^{110} - 2E^{101}),  \nonumber  \\
 K_{4,4} & = & \frac{1}{2} (E^{100} - E^{110}).       \nonumber
%%%% for (\sin^4(\phi)+\cos^4(\phi)) NOT CHECKED !!!!!!!!!!!!!!!!!!!!!!!!!
% K_{2,0} & = & (E^{101} - E^{001}) \frac{(z^2 + 1)^2}{z^2}, \nonumber  \\
% K_{4,0} & = & (E^{101} - E^{001}) (z^2 + 1)                \nonumber  \\
%         &   & +(E^{101} - E^{100}) \frac{(z^2 + 1)}{z^2},  \nonumber  \\
% K_{4,4} & = & 2 (E^{100} - E^{110}).                       \nonumber
\end{eqnarray}

The magnetocrystalline anisotropy energy ($E_{a}$) is the difference between
the magnetocrystalline energy (here $E^{\rm tet})$ and the isotropic
contribution, which is the spherical part $K_0$:

\begin{equation}
  E_{a} = E^{\rm tet} - K_0.
\label{eq:ea}
\end{equation}

%%%%%%%%%%%%%%%%%%%%%%%%%%%%%%%%%%%%%%%%%%%%%%%%%%%%%%%%%%%%%%%%%%%%%%%%%%%%%%%
\subsection{Dipolar magnetic anisotropy}
\label{app:app1_dipani}

In non-cubic systems, the dipolar anisotropy does not vanish
and also contributes to the magnetocrystalline anisotropy.
It is calculated from a direct
lattice sum yielding the dipolar energy:

\begin{equation}
\label{eq:dipaniso}
   E_{dip} (\vec{n}) = \frac{\mu_0}{8\pi} \sum_{i\neq j} \left[ \frac{\vec{m}_i \cdot \vec{m}_j}{r_{ij}^3}
                       - 3\frac{(\vec{r}_{ij}\cdot \vec{m}_i)(\vec{r}_{ij}\cdot \vec{m}_j)} {r_{ij}^5} \right],
\end{equation}

where $\vec{n}=\vec{M}/M$ is the magnetisation direction, and $r_{ij}$
represents the distance vectors between the magnetic moments $m_i$ and $m_j$.
The individual magnetic moments, $\vec{m}_i$ and $\vec{m}_j$, do not necessarily
have to be collinear in general.

In a simplified picture, only the $3d$ transition elements $T$ carry a
significant magnetic moment in the Rh$_2T$Sb compounds investigated here. In all
cases of a single magnetic ion where all the magnetic moments in the structure
are collinear along $\vec{n}$, the equation simplifies to

\begin{eqnarray}
\label{eq:dipsimple}
   E_{dip} (\vec{n}) & = & \frac{\mu_0 m^2(\vec{n})}{8\pi} \sum_{i\neq j} \frac{1}{r_{ij}^3} \left[ 1
                           - 3\frac{r^2_{n,ij}} {r_{ij}^2} \right], \nonumber \\
                     & = & \frac{\mu_0 m^2(\vec{n})}{8\pi} \sum_{i\neq j} \frac{1-3\cos^2(\theta_{ij})}{r_{ij}^3},
\end{eqnarray}

where $r_{n,ij}=r_{n,ij}(\vec{n})$ is a projection of the position vector onto
the direction of the magnetic moment, and $\theta_{ij}$ is the angle between
them. In Equation~(\ref{eq:dipsimple}), the sign of the energy is completely
defined by the crystal structure when the summation is over a spherical
particle. Note that the size of the magnetic moment, $m(\vec{n})$, depends on
the magnetisation direction when the spin--orbit interaction is taken into
account.

Finally, the dipolar anisotropy is given by the difference between the energies
for two different directions,
$\Delta E_{\rm dipaniso}=E(\vec{n}_2)- E(\vec{n}_1)$. Again, the two well-
distinguished directions are the $\vec{n}_1=[001]$ and $\vec{n}_2=[100]$
directions, which are along the $c$ axis and in the basal plane along $a$,
respectively. Positive values indicate an easy dipolar direction that is along
the $[001]$ axis. It has a second-order angular dependence.

%%%%%%%%%%%%%%%%%%%%%%%%%%%%%%%%%%%%%%%%%%%%%%%%%%%%%%%%%%%%%%%%%%%%%%%%%%%%%%%
\subsection{Plotting the magnetic anisotropy}
\label{app:app1_plot}

According to Equations~(\ref{eq:eu}) and~(\ref{eq:ea}), the magnetocrystalline
anisotropy energy may be positive or negative, depending on the direction of
($\theta,\phi$) and the $K$ values. Consequently, it is difficult to visualise
the anisotropy energy by plotting the three-dimensional distribution of
$E_a(\vec{r})=E_a(\theta,\phi)$. Therefore, the alternative anisotropy energy
$E_{a'}$ with respect to the lowest energy is generally plotted, where

\begin{equation}
  E_{a'} = E_{a} - \min(E_{a}),
\label{eq:eap}
\end{equation}

which is still positive even when $E_a<0$. The easy directions or planes are
identified as those for which $E_{a'}=0$. $E_{a'}$ is used to plot the
magnetocrystalline anisotropy in the main text.

%%%%%%%%%%%%%%%%%%%%%%%%%%%%%%%%%%%%%%%%%%%%%%%%%%%%%%%%%%%%%%%%%%%%%%%%%%%%%%%%%
\bigskip
%%%%%%%%%%%%%%%%%%%%%%%%%%%%%%%%%%%%%%%%%%%%%%%%%%%%%%%%%%%%%%%%%%%%%%%%%%%%%%%%
\begin{acknowledgments}

We thank the groups of P. Blaha (Vienna) and H. Ebert (Munich) for providing their
computer codes.

\end{acknowledgments}
%%%%%%%%%%%%%%%%%%%%%%%%%%%%%%%%%%%%%%%%%%%%%%%%%%%%%%%%%%%%%%%%%%%%%%%%%%%%%%%%

%%%%%%%%%%%%%%%%%%%%%%%%%%%%%%%%%%%%%%%%%%%%%%%%%%%%%%%%%%%%%%%%%%%%%%%%%%%%%%%%%
\bigskip
%%%%%%%%%%%%%%%%%%%%%%%%%%%%%%%%%%%%%%%%%%%%%%%%%%%%%%%%%%%%%%%%%%%%%%%%%%%%%%%%%
\bibliography{Rh2CoSb}   % Produces the bibliography via BibTeX.

\begin{thebibliography}{28}%
\makeatletter
\providecommand \@ifxundefined [1]{%
 \@ifx{#1\undefined}
}%
\providecommand \@ifnum [1]{%
 \ifnum #1\expandafter \@firstoftwo
 \else \expandafter \@secondoftwo
 \fi
}%
\providecommand \@ifx [1]{%
 \ifx #1\expandafter \@firstoftwo
 \else \expandafter \@secondoftwo
 \fi
}%
\providecommand \natexlab [1]{#1}%
\providecommand \enquote  [1]{``#1''}%
\providecommand \bibnamefont  [1]{#1}%
\providecommand \bibfnamefont [1]{#1}%
\providecommand \citenamefont [1]{#1}%
\providecommand \href@noop [0]{\@secondoftwo}%
\providecommand \href [0]{\begingroup \@sanitize@url \@href}%
\providecommand \@href[1]{\@@startlink{#1}\@@href}%
\providecommand \@@href[1]{\endgroup#1\@@endlink}%
\providecommand \@sanitize@url [0]{\catcode `\\12\catcode `\$12\catcode
  `\&12\catcode `\#12\catcode `\^12\catcode `\_12\catcode `\%12\relax}%
\providecommand \@@startlink[1]{}%
\providecommand \@@endlink[0]{}%
\providecommand \url  [0]{\begingroup\@sanitize@url \@url }%
\providecommand \@url [1]{\endgroup\@href {#1}{\urlprefix }}%
\providecommand \urlprefix  [0]{URL }%
\providecommand \Eprint [0]{\href }%
\providecommand \doibase [0]{https://doi.org/}%
\providecommand \selectlanguage [0]{\@gobble}%
\providecommand \bibinfo  [0]{\@secondoftwo}%
\providecommand \bibfield  [0]{\@secondoftwo}%
\providecommand \translation [1]{[#1]}%
\providecommand \BibitemOpen [0]{}%
\providecommand \bibitemStop [0]{}%
\providecommand \bibitemNoStop [0]{.\EOS\space}%
\providecommand \EOS [0]{\spacefactor3000\relax}%
\providecommand \BibitemShut  [1]{\csname bibitem#1\endcsname}%
\let\auto@bib@innerbib\@empty
%</preamble>
\bibitem [{\citenamefont {Dhar}\ \emph {et~al.}(1980)\citenamefont {Dhar},
  \citenamefont {Grover}, \citenamefont {Malik},\ and\ \citenamefont
  {Vijayaraghavan}}]{DGM80}%
  \BibitemOpen
  \bibfield  {author} {\bibinfo {author} {\bibfnamefont {S.~K.}\ \bibnamefont
  {Dhar}}, \bibinfo {author} {\bibfnamefont {A.~K.}\ \bibnamefont {Grover}},
  \bibinfo {author} {\bibfnamefont {S.~K.}\ \bibnamefont {Malik}},\ and\
  \bibinfo {author} {\bibfnamefont {R.}~\bibnamefont {Vijayaraghavan}},\
  }\bibfield  {title} {\bibinfo {title} {{Peaks in low field a.c.
  susceptibility of ferromagnetic Heusler alloys}},\ }\href@noop {} {\bibfield
  {journal} {\bibinfo  {journal} {Sol. St. Comm.}\ }\textbf {\bibinfo {volume}
  {33}},\ \bibinfo {pages} {545} (\bibinfo {year} {1980})}\BibitemShut
  {NoStop}%
\bibitem [{\citenamefont {Faleev}\ \emph {et~al.}(2017)\citenamefont {Faleev},
  \citenamefont {Ferrante}, \citenamefont {Jeong}, \citenamefont {Samant},
  \citenamefont {Jones},\ and\ \citenamefont {Parkin}}]{FFJ17}%
  \BibitemOpen
  \bibfield  {author} {\bibinfo {author} {\bibfnamefont {S.~V.}\ \bibnamefont
  {Faleev}}, \bibinfo {author} {\bibfnamefont {Y.}~\bibnamefont {Ferrante}},
  \bibinfo {author} {\bibfnamefont {J.}~\bibnamefont {Jeong}}, \bibinfo
  {author} {\bibfnamefont {M.~G.}\ \bibnamefont {Samant}}, \bibinfo {author}
  {\bibfnamefont {B.}~\bibnamefont {Jones}},\ and\ \bibinfo {author}
  {\bibfnamefont {S.~S.~P.}\ \bibnamefont {Parkin}},\ }\bibfield  {title}
  {\bibinfo {title} {{Heusler compounds with perpendicular magnetic anisotropy
  and large tunneling magnetoresistance}},\ }\href@noop {} {\bibfield
  {journal} {\bibinfo  {journal} {Phys. Rev. Materials}\ }\textbf {\bibinfo
  {volume} {1}},\ \bibinfo {pages} {024402} (\bibinfo {year}
  {2017})}\BibitemShut {NoStop}%
\bibitem [{\citenamefont {Blaha}\ \emph {et~al.}(1990)\citenamefont {Blaha},
  \citenamefont {Schwarz}, \citenamefont {Sorantin},\ and\ \citenamefont
  {Trickey}}]{BSS90}%
  \BibitemOpen
  \bibfield  {author} {\bibinfo {author} {\bibfnamefont {P.}~\bibnamefont
  {Blaha}}, \bibinfo {author} {\bibfnamefont {K.}~\bibnamefont {Schwarz}},
  \bibinfo {author} {\bibfnamefont {P.}~\bibnamefont {Sorantin}},\ and\
  \bibinfo {author} {\bibfnamefont {S.~B.}\ \bibnamefont {Trickey}},\
  }\bibfield  {title} {\bibinfo {title} {{Full-potential, linearized augmented
  plane wave programs for crystalline systems}},\ }\href@noop {} {\bibfield
  {journal} {\bibinfo  {journal} {Comput. Phys. Commun.}\ }\textbf {\bibinfo
  {volume} {59}},\ \bibinfo {pages} {399} (\bibinfo {year} {1990})}\BibitemShut
  {NoStop}%
\bibitem [{\citenamefont {Schwarz}\ and\ \citenamefont {Blaha}(2003)}]{SBl02}%
  \BibitemOpen
  \bibfield  {author} {\bibinfo {author} {\bibfnamefont {K.}~\bibnamefont
  {Schwarz}}\ and\ \bibinfo {author} {\bibfnamefont {P.}~\bibnamefont
  {Blaha}},\ }\bibfield  {title} {\bibinfo {title} {{Solid state calculations
  using WIEN2k}},\ }\href@noop {} {\bibfield  {journal} {\bibinfo  {journal}
  {Comput. Mater. Sci.}\ }\textbf {\bibinfo {volume} {28}},\ \bibinfo {pages}
  {259} (\bibinfo {year} {2003})}\BibitemShut {NoStop}%
\bibitem [{\citenamefont {Blaha}\ \emph {et~al.}(2013)\citenamefont {Blaha},
  \citenamefont {Schwarz}, \citenamefont {Madsen}, \citenamefont {Kvasnicka},\
  and\ \citenamefont {Luitz}}]{BSM01}%
  \BibitemOpen
  \bibfield  {author} {\bibinfo {author} {\bibfnamefont {P.}~\bibnamefont
  {Blaha}}, \bibinfo {author} {\bibfnamefont {K.}~\bibnamefont {Schwarz}},
  \bibinfo {author} {\bibfnamefont {G.~K.~H.}\ \bibnamefont {Madsen}}, \bibinfo
  {author} {\bibfnamefont {D.}~\bibnamefont {Kvasnicka}},\ and\ \bibinfo
  {author} {\bibfnamefont {J.}~\bibnamefont {Luitz}},\ }\href@noop {} {\emph
  {\bibinfo {title} {WIEN2k: An Augmented PlaneWave + Local Orbitals Program
  for Calculating Crystal Properties}}}\ (\bibinfo {address} {Wien},\ \bibinfo
  {year} {2013})\BibitemShut {NoStop}%
\bibitem [{\citenamefont {Ebert}(1999)}]{Ebe99}%
  \BibitemOpen
  \bibfield  {author} {\bibinfo {author} {\bibfnamefont {H.}~\bibnamefont
  {Ebert}},\ }\bibinfo {title} {Fully relativistic band structure calculations
  for magnetic solids - formalism and application},\ in\ \href@noop {} {\emph
  {\bibinfo {booktitle} {Electronic Structure and Physical Properties of
  Solids. The Use of the LMTO Method}}},\ \bibinfo {series} {Lecture Notes in
  Physics}, Vol.\ \bibinfo {volume} {535},\ \bibinfo {editor} {edited by\
  \bibinfo {editor} {\bibfnamefont {H.}~\bibnamefont {Dreysse}}}\ (\bibinfo
  {publisher} {Springer-Verlag},\ \bibinfo {address} {Berlin, Heidelberg},\
  \bibinfo {year} {1999})\ pp.\ \bibinfo {pages} {191 -- 246}\BibitemShut
  {NoStop}%
\bibitem [{\citenamefont {Ebert}\ \emph {et~al.}(2011)\citenamefont {Ebert},
  \citenamefont {K{\"o}dderitzsch},\ and\ \citenamefont {Minar}}]{EKM11}%
  \BibitemOpen
  \bibfield  {author} {\bibinfo {author} {\bibfnamefont {H.}~\bibnamefont
  {Ebert}}, \bibinfo {author} {\bibfnamefont {D.}~\bibnamefont
  {K{\"o}dderitzsch}},\ and\ \bibinfo {author} {\bibfnamefont {J.}~\bibnamefont
  {Minar}},\ }\bibfield  {title} {\bibinfo {title} {{Calculating condensed
  matter properties using the KKR-Green’s function method -- recent
  developments and applications}},\ }\href {stacks.iop.org/RoPP/74/096501}
  {\bibfield  {journal} {\bibinfo  {journal} {Rep. Prog. Phys}\ }\textbf
  {\bibinfo {volume} {74}},\ \bibinfo {pages} {096501} (\bibinfo {year}
  {2011})}\BibitemShut {NoStop}%
\bibitem [{\citenamefont {Perdew}\ \emph {et~al.}(1996)\citenamefont {Perdew},
  \citenamefont {Burke},\ and\ \citenamefont {Ernzerhof}}]{PBE96}%
  \BibitemOpen
  \bibfield  {author} {\bibinfo {author} {\bibfnamefont {J.~P.}\ \bibnamefont
  {Perdew}}, \bibinfo {author} {\bibfnamefont {K.}~\bibnamefont {Burke}},\ and\
  \bibinfo {author} {\bibfnamefont {M.}~\bibnamefont {Ernzerhof}},\ }\bibfield
  {title} {\bibinfo {title} {{Generalized Gradient Approximation Made
  Simple}},\ }\href@noop {} {\bibfield  {journal} {\bibinfo  {journal} {Phys.
  Rev. Lett.}\ }\textbf {\bibinfo {volume} {77}},\ \bibinfo {pages} {3865}
  (\bibinfo {year} {1996})}\BibitemShut {NoStop}%
\bibitem [{\citenamefont {Kandpal}\ \emph {et~al.}(2007)\citenamefont
  {Kandpal}, \citenamefont {Fecher},\ and\ \citenamefont {Felser}}]{KFF07b}%
  \BibitemOpen
  \bibfield  {author} {\bibinfo {author} {\bibfnamefont {H.~C.}\ \bibnamefont
  {Kandpal}}, \bibinfo {author} {\bibfnamefont {G.~H.}\ \bibnamefont
  {Fecher}},\ and\ \bibinfo {author} {\bibfnamefont {C.}~\bibnamefont
  {Felser}},\ }\bibfield  {title} {\bibinfo {title} {{Calculated electronic and
  magnetic properties of the half-metallic, transition metal based Heusler
  compounds}},\ }\href@noop {} {\bibfield  {journal} {\bibinfo  {journal} {J.
  Phys. D: Appl. Phys.}\ }\textbf {\bibinfo {volume} {40}},\ \bibinfo {pages}
  {1507 } (\bibinfo {year} {2007})}\BibitemShut {NoStop}%
\bibitem [{\citenamefont {Fecher}\ \emph {et~al.}(2013)\citenamefont {Fecher},
  \citenamefont {Chadov},\ and\ \citenamefont {Felser}}]{FCF13}%
  \BibitemOpen
  \bibfield  {author} {\bibinfo {author} {\bibfnamefont {G.~H.}\ \bibnamefont
  {Fecher}}, \bibinfo {author} {\bibfnamefont {S.}~\bibnamefont {Chadov}},\
  and\ \bibinfo {author} {\bibfnamefont {C.}~\bibnamefont {Felser}},\ }\bibinfo
  {title} {Theory of the half-metallic heusler compounds},\ in\ \href@noop {}
  {\emph {\bibinfo {booktitle} {Spintronics}}},\ \bibinfo {editor} {edited by\
  \bibinfo {editor} {\bibfnamefont {C.}~\bibnamefont {Felser}}\ and\ \bibinfo
  {editor} {\bibfnamefont {G.~H.}\ \bibnamefont {Fecher}}}\ (\bibinfo
  {publisher} {Springer Verlag},\ \bibinfo {address} {Dordrecht Heidelberg New
  York London},\ \bibinfo {year} {2013})\ \bibinfo {type} {Book
  section}~\bibinfo {chapter} {7}, p.\ \bibinfo {pages} {115}\BibitemShut
  {NoStop}%
\bibitem [{\citenamefont {Mankovsky}\ \emph {et~al.}(2011)\citenamefont
  {Mankovsky}, \citenamefont {Fecher},\ and\ \citenamefont {Ebert}}]{MFE11}%
  \BibitemOpen
  \bibfield  {author} {\bibinfo {author} {\bibfnamefont {S.}~\bibnamefont
  {Mankovsky}}, \bibinfo {author} {\bibfnamefont {G.~H.}\ \bibnamefont
  {Fecher}},\ and\ \bibinfo {author} {\bibfnamefont {H.}~\bibnamefont
  {Ebert}},\ }\bibfield  {title} {\bibinfo {title} {{Electronic structure
  calculations in ordered and disordered solids with spiral magnetic order}},\
  }\href@noop {} {\bibfield  {journal} {\bibinfo  {journal} {Phys. Rev. B}\
  }\textbf {\bibinfo {volume} {83}},\ \bibinfo {pages} {144401} (\bibinfo
  {year} {2011})}\BibitemShut {NoStop}%
\bibitem [{\citenamefont {Thoene}\ \emph {et~al.}(2009)\citenamefont {Thoene},
  \citenamefont {Chadov}, \citenamefont {Fecher}, \citenamefont {Felser},\ and\
  \citenamefont {K{\"u}bler}}]{TCF09}%
  \BibitemOpen
  \bibfield  {author} {\bibinfo {author} {\bibfnamefont {J.}~\bibnamefont
  {Thoene}}, \bibinfo {author} {\bibfnamefont {S.}~\bibnamefont {Chadov}},
  \bibinfo {author} {\bibfnamefont {G.}~\bibnamefont {Fecher}}, \bibinfo
  {author} {\bibfnamefont {C.}~\bibnamefont {Felser}},\ and\ \bibinfo {author}
  {\bibfnamefont {J.}~\bibnamefont {K{\"u}bler}},\ }\bibfield  {title}
  {\bibinfo {title} {{Exchange energies, Curie temperatures and magnons in
  Heusler compounds}},\ }\href@noop {} {\bibfield  {journal} {\bibinfo
  {journal} {J Phys. D: Appl. Phys.}\ }\textbf {\bibinfo {volume} {42}},\
  \bibinfo {pages} {084013} (\bibinfo {year} {2009})}\BibitemShut {NoStop}%
\bibitem [{\citenamefont {Soven}(1967)}]{Sov67}%
  \BibitemOpen
  \bibfield  {author} {\bibinfo {author} {\bibfnamefont {P.}~\bibnamefont
  {Soven}},\ }\bibfield  {title} {\bibinfo {title} {{Coherent-Potential Model
  of Substitutional Disordered Alloys}},\ }\href@noop {} {\bibfield  {journal}
  {\bibinfo  {journal} {Phys. Rev.}\ }\textbf {\bibinfo {volume} {156}},\
  \bibinfo {pages} {809} (\bibinfo {year} {1967})}\BibitemShut {NoStop}%
\bibitem [{\citenamefont {Khan}\ \emph {et~al.}(2016)\citenamefont {Khan},
  \citenamefont {Blaha}, \citenamefont {Ebert}, \citenamefont {Minar},\ and\
  \citenamefont {Sipr}}]{KBE16}%
  \BibitemOpen
  \bibfield  {author} {\bibinfo {author} {\bibfnamefont {S.~A.}\ \bibnamefont
  {Khan}}, \bibinfo {author} {\bibfnamefont {P.}~\bibnamefont {Blaha}},
  \bibinfo {author} {\bibfnamefont {H.}~\bibnamefont {Ebert}}, \bibinfo
  {author} {\bibfnamefont {J.}~\bibnamefont {Minar}},\ and\ \bibinfo {author}
  {\bibfnamefont {O.}~\bibnamefont {Sipr}},\ }\bibfield  {title} {\bibinfo
  {title} {{Magnetocrystalline anisotropy of FePt: A detailed view}},\ }\href
  {https://doi.org/10.1103/PhysRevB.94.144436} {\bibfield  {journal} {\bibinfo
  {journal} {Phys. Rev. B}\ }\textbf {\bibinfo {volume} {94}},\ \bibinfo
  {pages} {144436} (\bibinfo {year} {2016})}\BibitemShut {NoStop}%
\bibitem [{\citenamefont {Joshua}(1991)}]{Jos91}%
  \BibitemOpen
  \bibfield  {author} {\bibinfo {author} {\bibfnamefont {S.~J.}\ \bibnamefont
  {Joshua}},\ }\href@noop {} {\emph {\bibinfo {title} {Symmetry principles and
  magnetic symmetry in solid state physics.}}}\ (\bibinfo  {publisher} {Adam
  Hilger, IOP Publishing Ltd.},\ \bibinfo {address} {Bistol, Philadelphia, New
  York},\ \bibinfo {year} {1991})\BibitemShut {NoStop}%
\bibitem [{\citenamefont {Lizarraga}\ \emph {et~al.}(2004)\citenamefont
  {Lizarraga}, \citenamefont {Nordstr{\"o}m}, \citenamefont {Bergqvist},
  \citenamefont {Bergman}, \citenamefont {Sj{\"o}stedt}, \citenamefont {Mohn},\
  and\ \citenamefont {Eriksson}}]{LNB04}%
  \BibitemOpen
  \bibfield  {author} {\bibinfo {author} {\bibfnamefont {R.}~\bibnamefont
  {Lizarraga}}, \bibinfo {author} {\bibfnamefont {L.}~\bibnamefont
  {Nordstr{\"o}m}}, \bibinfo {author} {\bibfnamefont {L.}~\bibnamefont
  {Bergqvist}}, \bibinfo {author} {\bibfnamefont {A.}~\bibnamefont {Bergman}},
  \bibinfo {author} {\bibfnamefont {E.}~\bibnamefont {Sj{\"o}stedt}}, \bibinfo
  {author} {\bibfnamefont {P.}~\bibnamefont {Mohn}},\ and\ \bibinfo {author}
  {\bibfnamefont {O.}~\bibnamefont {Eriksson}},\ }\bibfield  {title} {\bibinfo
  {title} {{Conditions for Noncollinear Instabilities of Ferromagnetic
  Materials}},\ }\href@noop {} {\bibfield  {journal} {\bibinfo  {journal}
  {Phys. Rev. Lett.}\ }\textbf {\bibinfo {volume} {93}},\ \bibinfo {pages}
  {107205} (\bibinfo {year} {2004})}\BibitemShut {NoStop}%
\bibitem [{\citenamefont {Liechtenstein}\ \emph {et~al.}(1984)\citenamefont
  {Liechtenstein}, \citenamefont {Katsnelson},\ and\ \citenamefont
  {Gubanov}}]{LKG84}%
  \BibitemOpen
  \bibfield  {author} {\bibinfo {author} {\bibfnamefont {A.~I.}\ \bibnamefont
  {Liechtenstein}}, \bibinfo {author} {\bibfnamefont {M.~I.}\ \bibnamefont
  {Katsnelson}},\ and\ \bibinfo {author} {\bibfnamefont {V.~A.}\ \bibnamefont
  {Gubanov}},\ }\bibfield  {title} {\bibinfo {title} {{Exchange interactions
  and spin-wave stiffness in ferromagnetic metals}},\ }\href@noop {} {\bibfield
   {journal} {\bibinfo  {journal} {J. Phys. F: Met. Phys.}\ }\textbf {\bibinfo
  {volume} {14}},\ \bibinfo {pages} {L125} (\bibinfo {year}
  {1984})}\BibitemShut {NoStop}%
\bibitem [{\citenamefont {Liechtenstein}\ \emph {et~al.}(1987)\citenamefont
  {Liechtenstein}, \citenamefont {Katsnelson}, \citenamefont {Antropov},\ and\
  \citenamefont {Gubanov}}]{LKA87}%
  \BibitemOpen
  \bibfield  {author} {\bibinfo {author} {\bibfnamefont {A.~I.}\ \bibnamefont
  {Liechtenstein}}, \bibinfo {author} {\bibfnamefont {M.~I.}\ \bibnamefont
  {Katsnelson}}, \bibinfo {author} {\bibfnamefont {V.~P.}\ \bibnamefont
  {Antropov}},\ and\ \bibinfo {author} {\bibfnamefont {V.~A.}\ \bibnamefont
  {Gubanov}},\ }\bibfield  {title} {\bibinfo {title} {{Local Spin Density
  Functional Approach to the Theory of Exchange Interactions in Ferromagnetic
  Metals and Alloys}},\ }\href@noop {} {\bibfield  {journal} {\bibinfo
  {journal} {J. Magn. Magn. Mater.}\ }\textbf {\bibinfo {volume} {67}},\
  \bibinfo {pages} {65} (\bibinfo {year} {1987})}\BibitemShut {NoStop}%
\bibitem [{\citenamefont {Pajda}\ \emph {et~al.}(2001)\citenamefont {Pajda},
  \citenamefont {Kudrnovsky}, \citenamefont {Turek}, \citenamefont {Drchal},\
  and\ \citenamefont {Bruno}}]{PKT01}%
  \BibitemOpen
  \bibfield  {author} {\bibinfo {author} {\bibfnamefont {M.}~\bibnamefont
  {Pajda}}, \bibinfo {author} {\bibfnamefont {J.}~\bibnamefont {Kudrnovsky}},
  \bibinfo {author} {\bibfnamefont {I.}~\bibnamefont {Turek}}, \bibinfo
  {author} {\bibfnamefont {V.}~\bibnamefont {Drchal}},\ and\ \bibinfo {author}
  {\bibfnamefont {P.}~\bibnamefont {Bruno}},\ }\bibfield  {title} {\bibinfo
  {title} {{Ab initio calculations of exchange interactions, spin-wave
  stiffness constants, and Curie temperatures of Fe, Co, and Ni}},\ }\href@noop
  {} {\bibfield  {journal} {\bibinfo  {journal} {Phys. Rev. B}\ }\textbf
  {\bibinfo {volume} {64}},\ \bibinfo {pages} {174402} (\bibinfo {year}
  {2001})}\BibitemShut {NoStop}%
\bibitem [{\citenamefont {Turek}\ \emph {et~al.}(2012)\citenamefont {Turek},
  \citenamefont {Kudrnovsky},\ and\ \citenamefont {Carva}}]{TKC12}%
  \BibitemOpen
  \bibfield  {author} {\bibinfo {author} {\bibfnamefont {I.}~\bibnamefont
  {Turek}}, \bibinfo {author} {\bibfnamefont {J.}~\bibnamefont {Kudrnovsky}},\
  and\ \bibinfo {author} {\bibfnamefont {K.}~\bibnamefont {Carva}},\ }\bibfield
   {title} {\bibinfo {title} {{Magnetic anisotropy energy of disordered
  tetragonal Fe-Co systems from ab initio alloy theory}},\ }\href
  {https://doi.org/10.1103/PhysRevB.86.174430} {\bibfield  {journal} {\bibinfo
  {journal} {Phys. Rev. B}\ }\textbf {\bibinfo {volume} {86}},\ \bibinfo
  {pages} {174430} (\bibinfo {year} {2012})}\BibitemShut {NoStop}%
\bibitem [{\citenamefont {Skomsky}\ and\ \citenamefont {Coey}(1999)}]{SCo99}%
  \BibitemOpen
  \bibfield  {author} {\bibinfo {author} {\bibfnamefont {R.}~\bibnamefont
  {Skomsky}}\ and\ \bibinfo {author} {\bibfnamefont {J.~M.~D.}\ \bibnamefont
  {Coey}},\ }\href@noop {} {\emph {\bibinfo {title} {Permanent Magnetism}}},\
  Studies in Condensed Matter Physics\ (\bibinfo  {publisher} {Taylor and
  Francis Group},\ \bibinfo {address} {New York},\ \bibinfo {year}
  {1999})\BibitemShut {NoStop}%
\bibitem [{\citenamefont {K{\"u}bler}(2000)}]{Kub00}%
  \BibitemOpen
  \bibfield  {author} {\bibinfo {author} {\bibfnamefont {J.}~\bibnamefont
  {K{\"u}bler}},\ }\href@noop {} {\emph {\bibinfo {title} {Theory of Itinerant
  Electron Magnetism}}}\ (\bibinfo  {publisher} {Clarendon Press},\ \bibinfo
  {address} {Oxford},\ \bibinfo {year} {2000})\BibitemShut {NoStop}%
\bibitem [{\citenamefont {Cullity}\ and\ \citenamefont {Graham}(2009)}]{CGr09}%
  \BibitemOpen
  \bibfield  {author} {\bibinfo {author} {\bibfnamefont {B.~D.}\ \bibnamefont
  {Cullity}}\ and\ \bibinfo {author} {\bibfnamefont {C.~D.}\ \bibnamefont
  {Graham}},\ }\href@noop {} {\emph {\bibinfo {title} {Introduction to Magnetic
  Materials; 2nd Ed.}}}\ (\bibinfo  {publisher} {John Wiley and Sons},\
  \bibinfo {address} {Hoboken},\ \bibinfo {year} {2009})\BibitemShut {NoStop}%
\bibitem [{\citenamefont {Coey}(2010)}]{Coe10}%
  \BibitemOpen
  \bibfield  {author} {\bibinfo {author} {\bibfnamefont {J.~M.~D.}\
  \bibnamefont {Coey}},\ }\href@noop {} {\emph {\bibinfo {title} {Magnetism and
  Magnetic Materials}}}\ (\bibinfo  {publisher} {Cambridge University Press},\
  \bibinfo {address} {Cambridge},\ \bibinfo {year} {2010})\BibitemShut
  {NoStop}%
\bibitem [{\citenamefont {Spaldin}(2011)}]{Spa11}%
  \BibitemOpen
  \bibfield  {author} {\bibinfo {author} {\bibfnamefont {N.~A.}\ \bibnamefont
  {Spaldin}},\ }\href@noop {} {\emph {\bibinfo {title} {Magnetic Materials, 2nd
  Ed. Fundamentals and Applications}}}\ (\bibinfo  {publisher} {Cambridge
  University Press},\ \bibinfo {address} {Cambridge},\ \bibinfo {year}
  {2011})\BibitemShut {NoStop}%
\bibitem [{\citenamefont {Krishnan}(2016)}]{Kri16}%
  \BibitemOpen
  \bibfield  {author} {\bibinfo {author} {\bibfnamefont {K.~M.}\ \bibnamefont
  {Krishnan}},\ }\href@noop {} {\emph {\bibinfo {title} {Fundamentals and
  Applications of Magnetic Materials}}}\ (\bibinfo  {publisher} {Oxord
  University Press},\ \bibinfo {address} {Oxord},\ \bibinfo {year}
  {2016})\BibitemShut {NoStop}%
\bibitem [{\citenamefont {Jensen}\ and\ \citenamefont
  {Bennemann}(2006)}]{JBe06}%
  \BibitemOpen
  \bibfield  {author} {\bibinfo {author} {\bibfnamefont {P.~J.}\ \bibnamefont
  {Jensen}}\ and\ \bibinfo {author} {\bibfnamefont {K.~H.}\ \bibnamefont
  {Bennemann}},\ }\bibfield  {title} {\bibinfo {title} {{Magnetic structure of
  films: Dependence on anisotropy and atomic morphology}},\ }\href
  {https://doi.org/j.surfrep.2006.02.001} {\bibfield  {journal} {\bibinfo
  {journal} {Srf. Sci Rep.}\ }\textbf {\bibinfo {volume} {61}},\ \bibinfo
  {pages} {129} (\bibinfo {year} {2006})}\BibitemShut {NoStop}%
\bibitem [{\citenamefont {Skomski}\ \emph {et~al.}(1998)\citenamefont
  {Skomski}, \citenamefont {Oepen},\ and\ \citenamefont {Kirschner}}]{SOK98}%
  \BibitemOpen
  \bibfield  {author} {\bibinfo {author} {\bibfnamefont {R.}~\bibnamefont
  {Skomski}}, \bibinfo {author} {\bibfnamefont {H.-P.}\ \bibnamefont {Oepen}},\
  and\ \bibinfo {author} {\bibfnamefont {J.}~\bibnamefont {Kirschner}},\
  }\bibfield  {title} {\bibinfo {title} {{Unidirectional anisotropy in
  ultrathin transition-metal films}},\ }\href
  {https://doi.org/10.1103/PhysRevB.58.11138} {\bibfield  {journal} {\bibinfo
  {journal} {Phys. Rev. B}\ }\textbf {\bibinfo {volume} {58}},\ \bibinfo
  {pages} {11138} (\bibinfo {year} {1998})}\BibitemShut {NoStop}%
\end{thebibliography}%

\end{document}